\documentclass[prc,reprint,a4paper,showpacs,amsmath]{revtex4-1}

\usepackage{graphicx,longtable}
\usepackage[colorlinks=true,allcolors=blue]{hyperref}

\LTcapwidth\textwidth\newcommand\empt{\phantom{000.000(0.000)}}

\begin{document}

\title{Nuclear masses near $N = Z$ from Nilsson-Strutinsky
  calculations with pairing corrections beyond BCS from an
  isospin-conserving pairing force}

\author{I. Bentley}
\affiliation{Department of Physics, University of Notre Dame,
  Notre Dame, Indiana 46556, USA,}
\affiliation{Department of Chemistry and Physics, Saint Mary's
  College, Notre Dame, Indiana 46556, USA}

\author{K. Neerg\aa rd}
\affiliation{Fjordtoften 17, 4700 N\ae stved, Denmark}

\author{S. Frauendorf}
\affiliation{Department of Physics, University of Notre Dame,
  Notre Dame, Indiana 46556, USA}

\begin{abstract}
  A model with nucleons in a charge-independent potential well
  interacting by an isovector pairing force is considered. For a
  24-dimensional valence space, the Hartree-Bogolyubov (HB) plus
  random phase approximation (RPA) to the lowest eigenvalue of the
  Hamiltonian is shown to be accurate except near values of the
  pairing-force coupling constant $G$ where the HB solution shifts
  from a zero to a nonzero pair gap. In the limit $G\to\infty$ the HB
  + RPA is asymptotically exact. The inaccuracy of the HB + RPA in the
  critical regions of $G$ can be remedied by interpolation. The
  resulting algorithm is used to calculate pairing corrections in the
  framework of a Nilsson-Strutinsky calculation of nuclear masses near
  $N=Z$ for $A=$~24--100, where $N$ and $Z$ are the numbers of
  neutrons and protons, and $A=N+Z$. The dimension of the valence
  space is 2$A$ in these calculations. Adjusting five liquid drop
  parameters and a power law expression for the constant $G$ as a
  function of $A$ allows us to reproduce the measured binding energies
  of 112 doubly even nuclei in this range with a root mean square
  deviation of 0.95~MeV. Several combinations of the masses for
  different $N$, $Z$, and isospin $T$ are considered and the
  calculations found to be in good agreement with the data. It is
  demonstrated by examples how fluctuations as a function of $A$ of
  the constant $X$ in an expansion of the symmetry energy of the form
  $T(T+X)/2\theta$ can be understood from the shell structure.
\end{abstract}


\maketitle

\section{\label{sec:Intro}Introduction}

Since the late 1990s the masses of nuclei near the line $N = Z$ in the
chart of nuclides, where $N$ and $Z$ are the numbers of neutrons and
protons, have attracted much interest from the nuclear physics
community. In particular, the origin of the so-called Wigner
energy~\cite{MySw}, a depression of the mass at $N = Z$ relative to a
trend described by a symmetry energy quadratic in $N - Z$, has been a
matter of debate. For a review of this discussion, see Neerg\aa
rd~\cite{Ne09}. In 1995, Duflo and Zuker~\cite{DZ95} published a
semiempirical mass formula with a symmetry energy proportional to
$T(T+1)$, where $T$ is the isospin quantum number, in the ground
states of doubly even nuclei equal to $|T_z|$, where $T_z=(N-Z)/2$.
This expression includes a Wigner energy in a natural manner. It was
observed by Frauendorf and Sheikh~\cite{FrSh} that a symmetry energy
with the factor $T(T+1)$ resembles the spectrum of a quantal, axially
symmetric rotor. These authors identified the nuclear superfluidity as
the deformation in isospace that could give rise to collective
rotation in this space. The Bardeen-Cooper-Schrieffer (BCS) pair gaps
$\Delta_{n,p}$ are indeed components of an isovector perpendicular to
the isospin. Their magnitude is a measure of the collectivity of the
isorotation. Obviously the dependence of the energy on $T$ will become
more regular as their magnitude increases as a result of the
progressively more gradual change of the occupation numbers around the
Fermi surface, which will wash out fluctuations of the level density.
On the other hand the isorotational moment of inertia, which is
determined by the average level density, will not change much. The
results of calculations in Ref. \cite{Sat00} exemplify these generic
features.

Neerg\aa rd~\cite{Ne02,Ne03,Ne09} set up a microscopic theory of such
a superfluid isorotation based on the Hartree-Bogolyubov (HB) plus
random phase approximation (RPA). His Hamiltonian involves independent
nucleons in a charge-independent potential well, an isovector pairing
force, and an interaction of the nucleonic isospins, which he calls
the symmetry force. The latter is shown to contribute merely a term in
the total energy proportional to $T(T+1)$. In the idealized case of
equidistant single-nucleon levels, the \emph{total} symmetry energy is
found to be proportional to $T(T+1)$ provided the pairing force is
sufficiently strong to produce an HB energy minimum with nonzero BCS
gaps. If this condition fails to be satisfied, the model gives in this
case a total symmetry energy proportional to $T(T+X)$ with
$X<1$~\cite{Ne03}. In realistic cases with nonuniform single-nucleon
spectra, major modifications of these simple expressions arise from
shell effects~\cite{Ne09}.

Recently, Bentley and Frauendorf~\cite{BF13} calculated exactly the
lowest eigenvalue of Neerg\aa rd's Hamiltonian employing small valence
spaces of dimensions 24 or 28. They demonstrated (see their Fig.~7)
that for a sufficiently strong pairing interaction the $T(T+1)$ limit
of rigid isorotation is approached for various kinds of bunched
single-particle level distributions. Unlike Neerg\aa rd, who keeps the
Hamiltonian constant along each isobaric chain, these authors take
into account the variation of the nuclear shape with the isospin. In a
survey of the range of mass numbers $A=N+Z=$ 24--100, they find that
their model accounts generally for the fluctuations due to the shell
structure observed in several combinations of the masses near $N=Z$
taken as functions of $A$: (i) the mass difference of doubly odd and
doubly even $N=Z$ nuclei, (ii) the difference in excitation energy of
the lowest $T=1$ and $T=0$ states in the the doubly odd $N=Z$ nuclei,
and (iii) the constants $\theta$ and $X$ in an expansion of the
symmetry energy of the form $T(T+X)/2\theta$ extracted from doubly
even masses close to $T=0$. However, the model underestimates $\theta$
when the symmetry force constant is fit to the difference of $T=1$ and
$T=0$ excitation energies.

The exact results of Bentley and Frauendorf provide a background on
which the accuracy of the HB + RPA may be tested. We show in the
present study that the HB + RPA gives a very good approximation to the
exact lowest eigenvalue of the Hamiltonian except near the values of
the pairing force coupling constant $G$ where the HB solution shifts
from a zero to a non-zero pair gap. We show, as well, that the HB +
RPA reproduces the exact eigenvalue asymptotically in the limit
$G\to\infty$. We then devise a recipe for interpolating the HB + RPA
energy across the critical region of $G$ leading to an algorithm which
accurately approximates the exact eigenvalue in the entire range of
$G$ from zero to infinity.

This algorithm is simple enough to allow calculations with valence
spaces of dimension $2A$. More specifically, we include in our present
calculations all single-nucleon states below the $T=0$ Fermi level and
equally many states above this level. To allow actual nuclear masses
to be calculated from this schematic model, we add a Strutinsky
renormalization. As a side effect, we can then dispense with the
symmetry force. Its contribution to the total energy proportional to
$T(T+1)$ may thus be considered a part of the macroscopic liquid drop
symmetry energy. Our microscopic Hamiltonian thus consists merely of a
charge-independent independent-nucleon term and the isovector pairing
force.

Using this scheme we calculate once more the combinations of masses
near $N=Z$ previously considered by Bentley and Frauendorf and show
that the issue with the underestimation of the constant $\theta$ is
resolved by the enlargement of the valence space. We also show that
the present scheme allows to calculate the masses themselves with a
small root-mean-square deviation from the measured ones.

The following Secs.~\ref{sec:BE}--\ref{sec:RPA} and~\ref{sec:Smooth}
contain the formal presentation of our model. In
Sec.~\ref{sec:Interpolation} we compare the exact lowest eigenstate of
our microscopic Hamiltonian with the HB + RPA and explain our scheme
of interpolation. Then, in Sec.~\ref{sec:Comparison}, we present and
analyze our results of calculation. Sec.~\ref{sec:NoIsoscalar}
provides arguments detailing why we disregard isoscalar pair
correlations. Finally, in Sec.~\ref{sec:Sum}, the article is
summarized and some perspectives drawn.

\section{\label{sec:BE}Liquid drop energy}

Following Strutinsky (Ref.~\cite{BDJ72} and references therein) we
assume that the nuclear binding energy $B(A,T,T_z)$ for mass number
$A$ and isospin $(T,T_z)$ is given by
\begin{equation}\label{eqn:Strut}
  - B(A,T,T_z) = E_\text{DLD} + E_\text{s.p.} - \tilde E_\text{s.p.}
                 + P - \tilde P \,.
\end{equation}
Here $E_\text{DLD}$ is a deformed liquid drop energy, $E_\text{s.p.}$
is the sum of occupied single-nucleon levels in a generally deformed
potential well, $P$ is the pairing energy, and $\tilde E_\text{s.p.}$
and $\tilde P$ are ``smooth'' counterparts of $E_\text{s.p.}$ and $P$.
For $E_\text{DLD}$ we adopt an expression of the form proposed by
Duflo and Zuker~\cite{DZ95},
\begin{equation}\label{eqn:LDM}\begin{aligned}
  E_\text{DLD} =
    & - \left(a_v - a_{vt} \frac{T(T+1)}{A^2} \right) A \\
    & +
      \left(a_s -a_{st}  \frac{T(T+1)}{A^2} \right) A^{2/3} B_S \\
    & + a_c \frac{Z(Z-1)}{A^{1/3}} B_C \,,
\end{aligned}\end{equation}
where $B_S$ and $B_C$ are functions of the shape. As explained below
we need these functions only for axial and reflection symmetry. We
then employ the expansions in the Hill-Wheeler deformation parameters
$\alpha_2$ and $\alpha_4$~\cite{HW53} given by Swiatecki~\cite{Sw56}
with $\alpha_2$ and $\alpha_4$ expressed in turn by the Nilsson
deformation parameters $\epsilon_2$ and $\epsilon_4$~\cite{Ni69} by
means of the expansions given by Seeger and Howard~\cite{SH75}. The
determination of $\epsilon_2$ and $\epsilon_4$ is discussed in
Sec.~\ref{sec:Hamiltonian}.

The five parameters in Eq.~\eqref{eqn:LDM} are determined by fitting
Eq.~\eqref{eqn:Strut} to the 112 measured binding energies of doubly
even nuclei considered in the present study according to the 2012
Atomic Mass Evaluation~\cite{Au12} with
$E_\text{s.p.}$, $\tilde E_\text{s.p.}$, $P$, and $\tilde P$
calculated as described in
Secs.~\ref{sec:Hamiltonian}--\ref{sec:Smooth}. The result is
$a_v = 15.19$~MeV, $a_{vt} =110.8$~MeV, $a_s = 16.35$~MeV,
$a_{st} = 135.3$~MeV, and $a_c = 0.6615$~MeV. These parameters are
similar to those obtained by Mendoza-Temis~\textit{et al.} in a global
fit of nuclear masses with minus the binding energy given by
Eq.~\eqref{eqn:LDM} plus a phenomenological, negative definite,
pairing energy and $B_S = B_C = 1$~\cite{MHZ10}. The parameters of
Mendoza-Temis~\textit{et~al.} cannot be used in the present context
because unlike their negative definite pairing energy the sum of the
liquid drop deformation energy and the shell and pairing correction
terms in Eq.~\eqref{eqn:Strut} average approximately to zero.
Moreover, the semiempirical formula of Mendoza-Temis~\textit{et~al.}
deviates quite a lot from the empirical binding energies locally in
the present region of nuclei. This deviation increases with increasing
$A$ along the $N = Z$ line and amounts to about 20~MeV for $A \approx
100$. Since our liquid drop parameters are optimized for the present
region of $N$ and $Z$ they may not reproduce accurately binding
energies in other regions.

\section{\label{sec:Hamiltonian}Single-particle plus pairing
  Hamiltonian}

The sum $E = E_\text{s.p.} + P$ in Eq.~\eqref{eqn:Strut} is calculated
as the lowest eigenvalue of the Hamiltonian
\begin{equation}\label{eqn:IVPH}
 H = \sum_{k} \epsilon_{k} \hat{N}_k
     - G \sum_{kk', \tau} \hat{P}^+_{k,\tau} \hat{P}_{k',\tau} \,,
\end{equation}
where $k$ labels orthogonal quartets of a single-proton and a
single-neutron state and their time reversed. The annihilator of a
nucleon in one of these states is denoted by $\hat p_k$, $\hat n_k$,
$\hat p_{\bar k}$, or $\hat n_{\bar k}$, and
\begin{equation}\begin{gathered}
  \hat N_{k} = \hat p^+_k \hat p_k
              + \hat p^+_{\bar k} \hat p_{\bar{k}}
              + \hat n^+_k \hat n_k
              + \hat n^+_{\bar k} \hat n_{\bar{k}} \,, \\
  \hat P^+_{k,0} = \frac1{\sqrt 2}\left(
                    \hat n^+_k \hat p^+_{\bar{k}}
                    + \hat p^+_{k} \hat n^+_{\bar{k}} \right) \,, \\
  \hat P^+_{k,-1} = \hat p^+_k \hat p^+_{\bar k} \,, \quad
  \hat P^+_{k,1} = \hat n^+_k \hat n^+_{\bar k} \,.
\end{gathered}\end{equation}
The single-nucleon energies $\epsilon_k$ are derived from a
calculation with the Nilsson potential employing the parameters of
Bengtsson and Ragnarsson~\cite{Beng85}. To conserve isospin we take
the average of the neutron and proton energies with given ordinal
number counted from the bottom of the spectrum. The resulting quartets
are labeled by $k$ in the order of increasing $\epsilon_k$, and the
first $\Omega = A/2$ of them in this order included in the
calculation.

The deformation parameters $\epsilon_2$ and $\epsilon_4$ are taken
from a recent survey of deformations based on the Nilsson-Strutinsky
plus BCS theory. The equilibrium deformation of a given nucleus was
calculated in this survey by minimizing with respect to Larsson's
triaxial deformation parameters $\epsilon$ and $\gamma$~\cite{La73} as
well as $\epsilon_4$ the Nilsson-Strutinsky plus BCS energy calculated
with the \textsc{tac} code~\cite{Fr93}. The latter employs an
expression for the liquid drop energy similar to Eq.~\eqref{eqn:LDM}
but with symmetry energy terms quadratic in $T_z$, the Nilsson
potential with the parameters of Ref.~\cite{Beng85}, and a pairing
correction without a smooth counterterm calculated from BCS gaps
$\Delta_n=\Delta_p=12A^{-1/2}$~MeV. In calculations for odd $N$ or $Z$
the Fermi level is blocked. All the 136 nuclei included in our study
turn out to have either $\gamma = 0$ or $\gamma = 60^\circ$, that is,
axial symmetry. The deformations are shown in Tables~\ref{tab:LDparam}
and~\ref{tab:LDparam4}. Also see Note~\cite{supa}.

The only free parameter remaining is the pairing force coupling
constant $G$. A power law for $G$ as a function of $A$ will be fit to
$T=0$ doubly even-doubly odd binding energy differences.

\section{\label{sec:RPA}Hartree-Bogolyubov plus Random Phase
  Approximation}
   
For an introduction to the BCS, HB, and RPA theories we refer to
textbooks such as the one by Ring and Schuck~\cite{RS80}. The
calculation of the lowest eigenvalue of the
Hamiltonian~\eqref{eqn:IVPH} in the HB + RPA for even $N$ and $Z$ is
discussed by Neerg\aa rd~\cite{Ne09}. Since this formalism is
invariant under isorotation, a nucleus with $N \ge Z$ represents the
entire multiplet with $T=T_z= (N - Z)/2$. To calculate the energy of
the lowest $T = 0$ state for odd $N = Z$ we reduce this case to the
even one by omitting quartet number $k = (N + 1)/2$ from a HB + RPA
calculation for $N - 1$, $Z - 1$ and adding $2\epsilon_k$.

The HB part of the calculation amounts to the usual BCS theory with
quasinucleons annihilators
\begin{equation}\begin{gathered}
  \hat\alpha_k = u_k \hat n_k - v_k \hat n_{\bar k}^+ \,, \quad
  \hat\alpha_{\bar k} = u_k \hat n_{\bar k} + v_k \hat n_k^+ \,, \\
  \hat\beta_k = w_k \hat p_k - z_k \hat p_{\bar k}^+ \,, \quad
  \hat\beta_{\bar k} = w_k \hat p_{\bar k} + z_k \hat p_k^+ \,.
\end{gathered}\end{equation}
For a gap $\Delta = 0$ the chemical potential $\lambda$ is taken as
the limit for $\Delta \to 0$ of the $\lambda$ determined by $N$ or $Z$
for $\Delta > 0$.

The RPA part splits into separate equations for an ``$n$ space''
spanned by $\hat\alpha_{\bar k}\hat\alpha_k$ and their Hermitian
conjugates, a ``$p$ space'' spanned by $\hat\beta_{\bar k}\hat\beta_k$
and their Hermitian conjugates, and an ``$np$ space'' spanned by
$\hat\beta_{\bar k}\hat\alpha_k + \hat\alpha_{\bar k}\hat\beta_k$ and
their Hermitian conjugates. The resulting groundstate energy can be
written as $E_\text{s.p.} + P$ with $P = P_\text{BCS} + P_\text{RPA}$
and
\begin{equation}\begin{gathered}
  E_\text{s.p.} = 2 \left( \sum_{k\le N/2} \epsilon_k
    + \sum_{k\le Z/2} \epsilon_k \right) \,, \\[5pt]
  P_\text{BCS} = 2 \sum_k (v_k^2 + z_k^2) \epsilon_k
    - (\Delta_n^2 + \Delta_p^2)/G - E_\text{s.p.} \,, \\
  P_\text{RPA}
    = E_\text{RPA,$n$} + E_\text{RPA,$p$} + E_\text{RPA,$np$} \,.
\end{gathered}\end{equation}
Here $\Delta_n$ and $\Delta_p$ are the neutron and proton BCS gaps and
\begin{equation}\label{eqn:RPA}\begin{gathered}
  E_\text{RPA,$n$ or $p$}
    = \tfrac12 \sum_\text{$n$ or $p$ space} \omega
      \ - \ \sum_k E_{k,n\text{ or }p} \,, \\
  E_\text{RPA,$np$}
    = \tfrac12 \left( \sum_\text{$np$ space} \omega
      \ - \ \sum_k (E_{k,n} + E_{k,p}) \right) \,,
\end{gathered}\end{equation}
where $\omega$ denotes an RPA frequency and $E_{k,n\text{ or }p}$ are
the single-quasinucleon energies. (A term $c$ given by Eq.~(35) of
Ref.~\cite{Ne09} vanishes in the present case of %
$A = 2\Omega$.)

For nuclei with $N = Z$, the BCS solutions are equal for neutrons and
protons and the $n$, $p$, and $np$ spaces have equal RPA spectra so
that $E_\text{RPA,$n$} = E_\text{RPA,$p$} = E_\text{RPA,$np$}$. If
$\Delta_n > 0$ the lowest RPA frequency in the $n$ space vanishes
because the Hamiltonian~\eqref{eqn:IVPH} commutes with $N$, and the
analogon of this statement holds for protons. For $N > Z$ the lowest
frequency in the $np$ space is equal to the difference $\lambda_n -
\lambda_p$ of the neutron and proton chemical potentials because the
Hamiltonian commutes with the components of the isospin perpendicular
to the $z$ direction.

\section{\label{sec:Interpolation}Comparison with the exact lowest
  eigenvalue of the Hamiltonian}

\begin{figure}
  \includegraphics[width=\columnwidth,trim=0 30 0 0]{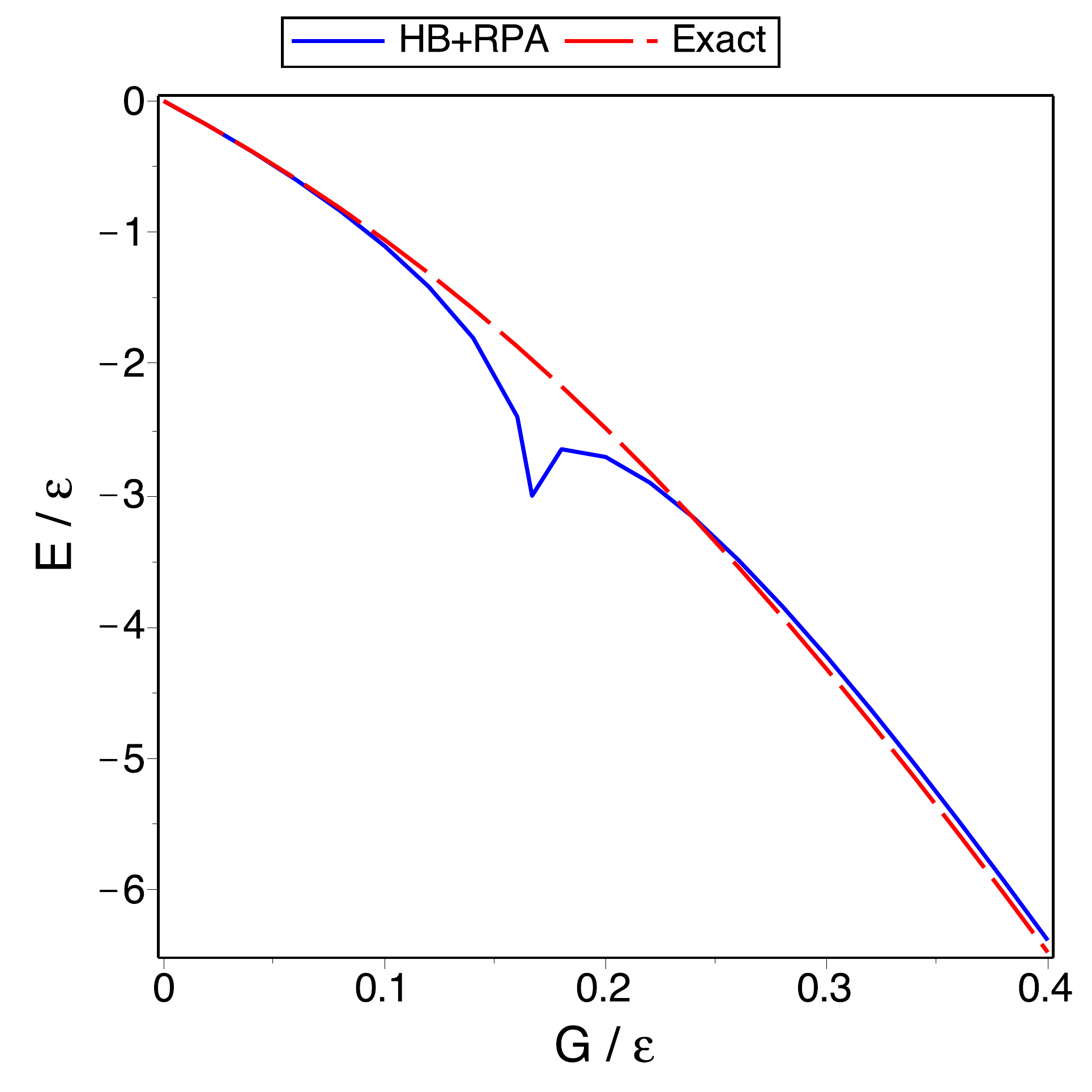}
  \caption{\label{fig:HBRPA-Exact}%
    (Color online) The energy $E$ calculated in the HB + RPA and
    exactly for 12-fold degenerate single-nucleon levels 0 and
    $\epsilon$ occupied for $G=0$ from the bottom by six neutrons and
    six protons.}
\end{figure}

Bentley and Frauendorf calculated exactly the lowest eigenvalue of the
Hamiltonian~\eqref{eqn:IVPH} in spaces with six or seven
quartets~\cite{BF13}. This allows a comparison of the HB + RPA to an
exact calculation. We made this comparison in all the cases displayed
in Fig.~7 of Ref.~\cite{BF13}. The case shown in
Fig.~\ref{fig:HBRPA-Exact}
is the one with the largest deviation of the HB + RPA result from the
exact energy. The HB + RPA curve is seen to follow closely the exact one
except in a small interval about $G = \epsilon/6$, which is in this
case the critical value $G_\text{crit.}$ of $G$ where the BCS solution
(equal in this case for neutrons and protons) changes from
$\Delta = 0$ to $\Delta > 0$. While the exact groundstate energy has
a smooth variation across $G_\text{crit.}$, the HB + RPA curve shows
there a prominent cusp.

\begin{figure}
  \includegraphics[width=\columnwidth,trim=0 30 0 0]{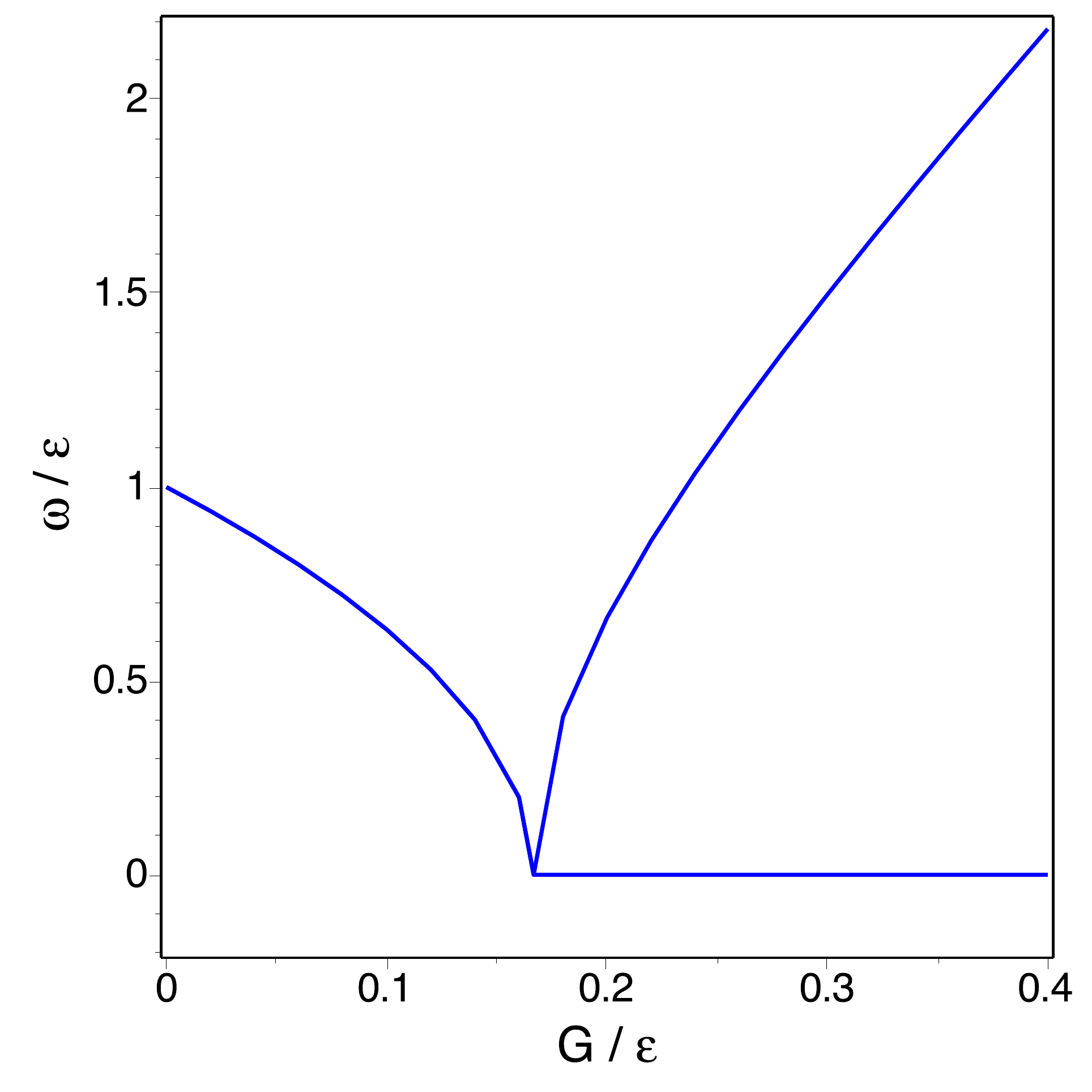}
  \caption{\label{fig:freq1}%
    (Color online) The two lowest RPA frequencies in any of the $n$,
    $p$, and $np$ spaces in the case of Fig.~\ref{fig:HBRPA-Exact}.
    These two frequencies coincide in the present case for %
    $G \le G_\text{crit.}$ because the single-nucleon spectrum is
    symmetric about the common Fermi level of neutrons and protons.}
\end{figure}

The origin of this cusp can be traced to the
expression~\eqref{eqn:RPA} for the RPA contributions. Thus notice the
plot in Fig.~\ref{fig:freq1}
of the two lowest RPA frequencies in the case just considered in any
of the $n$, $p$, and $np$ spaces, which have in this case identical
RPA spectra because $N = Z$. Both frequencies are seen to go to zero
for $G$ going to $G_\text{crit.}$ from below, but only one of them
stays at zero for $G > G_\text{crit.}$ while the other one rises
rapidly in this interval. The reason why two frequencies and not only
one go to zero for $G$ going to $G_\text{crit.}$ from below is that
the quasinucleon vacuum becomes in this limit instable against a
transition to a vacuum described by a non-zero $\Delta$ with an
arbitrary complex phase. This $\Delta$ has two real parameters. In
more physical terms, Fig.~\ref{fig:HBRPA-Exact} may be interpreted to
display a shortcoming of the RPA, which is a small amplitude
approximation, in a region of the parameter $G$ where the equilibrium
represented by the quasinucleon vacuum changes rapidly with this
parameter.

\begin{figure}
  \includegraphics[width=\columnwidth,trim=0 30 0 0]{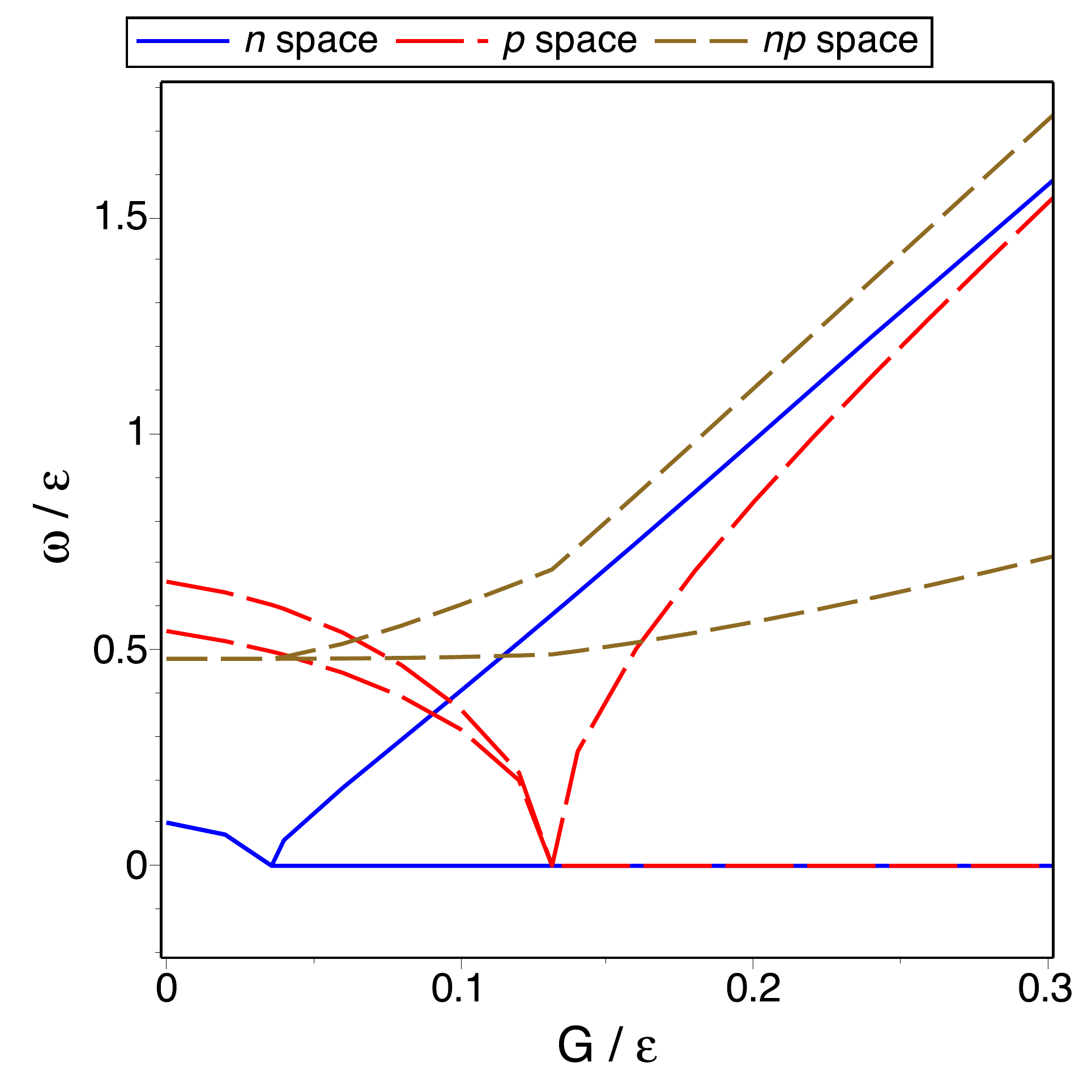}
  \caption{\label{fig:freq2}%
    (Color online) The two lowest RPA frequencies in each of the $n$,
    $p$, and $np$ spaces in the case of quartets at 0, $0.1\epsilon$,
    $0.7\epsilon$, $0.8\epsilon$, $0.9\epsilon$, and $\epsilon$
    occupied for $G = 0$ from the bottom by eight neutrons and four
    protons.}
\end{figure}

An example with less symmetry of the single-nucleon spectrum is given
in Fig.~\ref{fig:freq2}. %
It is seen that while the lowest RPA frequencies in the $n$ and $p$
spaces behave as in the preceding case, the lowest RPA frequencies in
the $np$ space pass almost smoothly through the critical $G$, which
are different in this case for neutrons and protons. This is general
for $N > Z$.

These observations suggest that the HB + RPA may be improved by
interpolation across the critical regions of $G$. More specifically we
have found that one gets a good approximation to the exact groundstate
energies by interpolation in the interval from $0.5\,G_\text{crit.}$
to $1.5\,G_\text{crit.}$. This is applied to the terms
$E_\text{RPA,$n$}$, $E_\text{RPA,$p$}$, and $E_\text{RPA,$np$}$ for $N
= Z$, when neutrons and protons have the same $G_\text{crit.}$, and to
$E_\text{RPA,$n$}$ and $E_\text{RPA,$p$}$ for $N > Z$, when
$G_\text{crit.}$ may be different for neutrons and protons. No
interpolation is applied if $G_\text{crit.} = 0$, which occurs when
the Fermi level lies within a degenerate shell. The interpolating
function is the polynomial of third degree in $G$ which joins smoothly
the calculated values at the endpoints of the interpolation interval.

\begin{figure}
  \includegraphics[width=\columnwidth,trim=0 30 0 0]{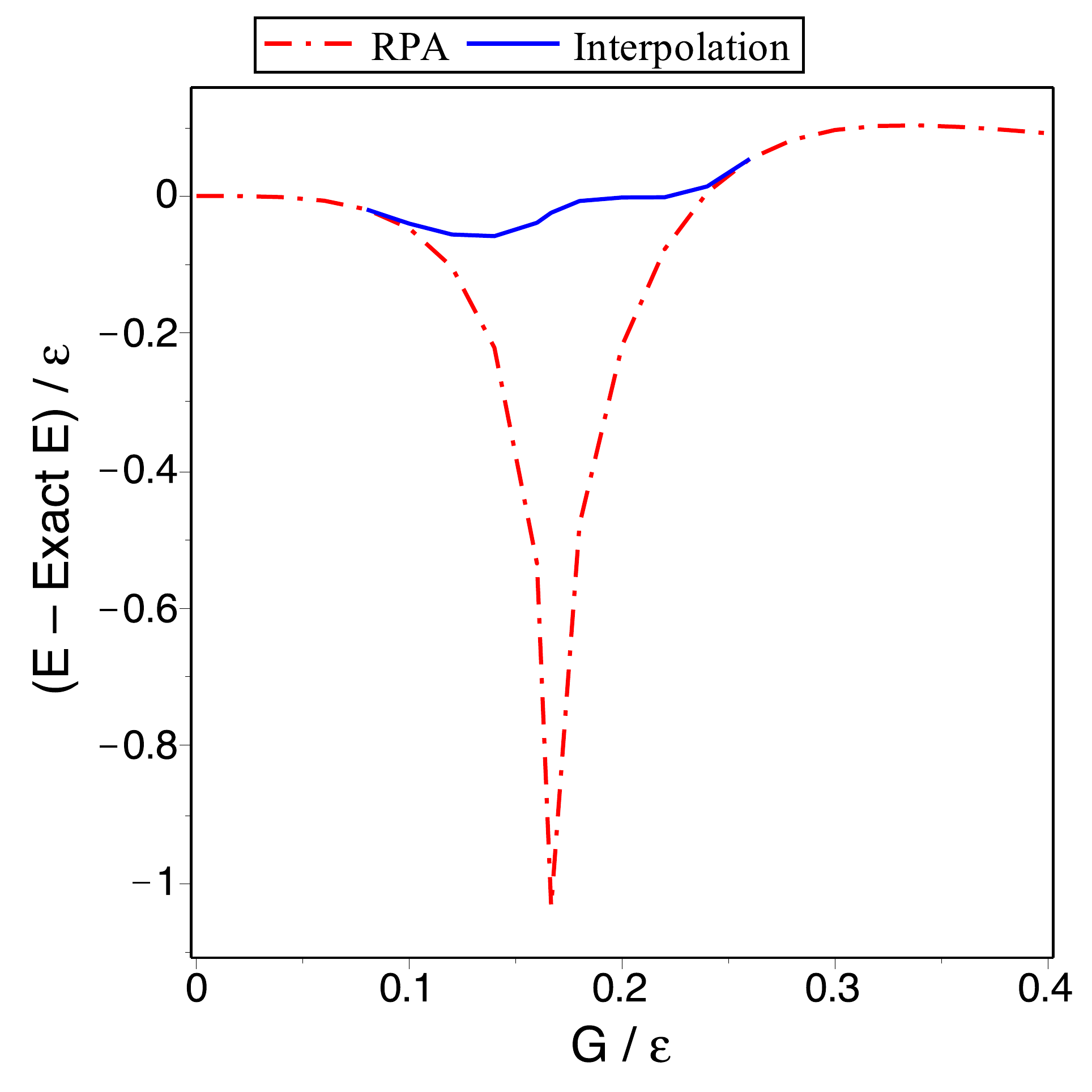}
  \caption{\label{fig:int1}%
    (Color online) The deviation of the HB + RPA result from the exact
    $E$ in the case of Fig. \ref{fig:HBRPA-Exact} without and with
    interpolation of the RPA part.}
\end{figure}

\begin{figure}
  \includegraphics[width=\columnwidth,trim=0 30 0 0]{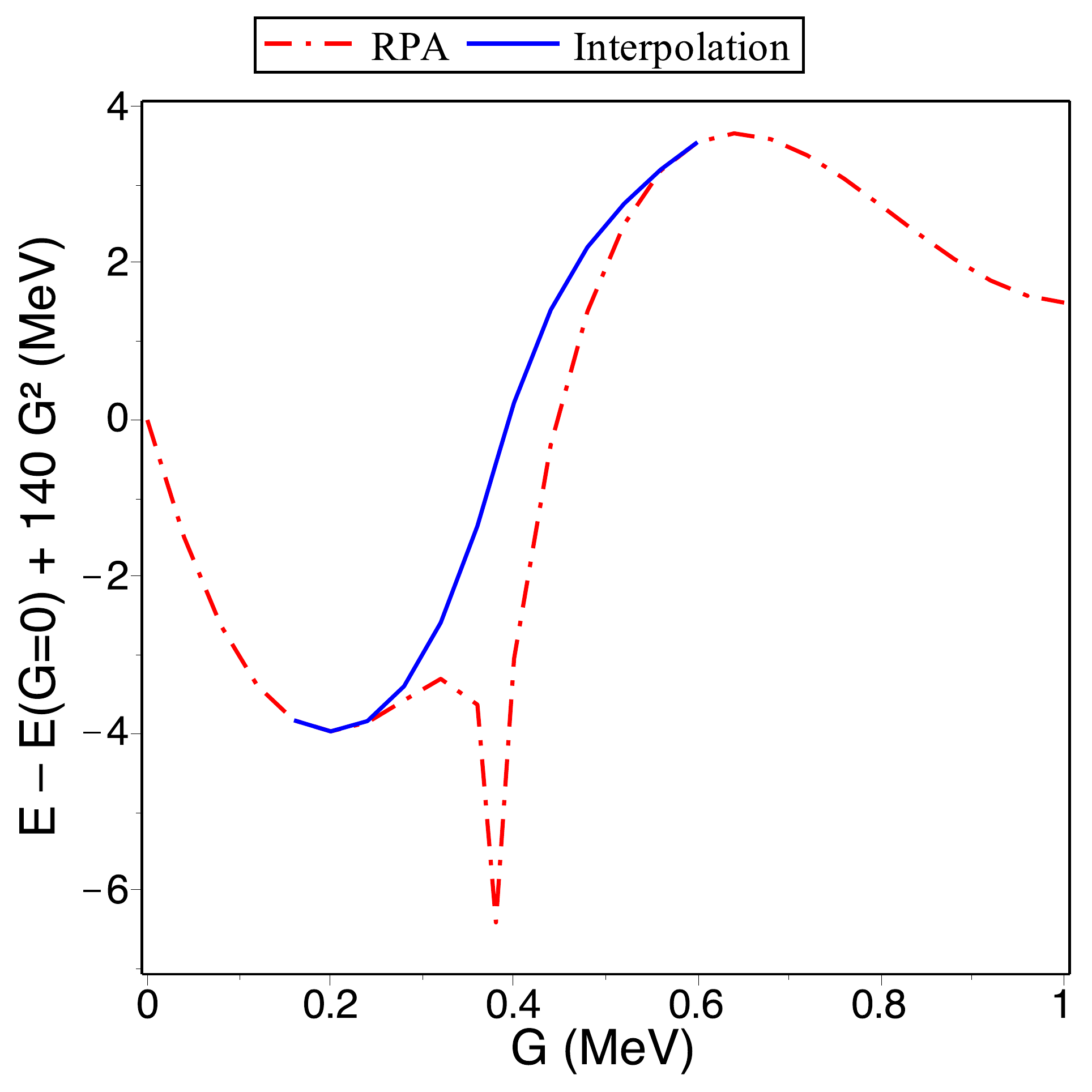}
  \caption{\label{fig:int2}%
    (Color online) The energy $E$ relative to $G = 0$ for the
    nucleus $^{56}$N calculated in the HB + RPA without and with
    interpolation of the RPA part. To enhance the details in the
    figure a term quadratic in $G$ is added to the calculated
    energies.}
\end{figure}

Figure~\ref{fig:int1}
shows the result of using this recipe in the ``worst case'' of
Fig.~\ref{fig:HBRPA-Exact}, and Fig.~\ref{fig:int2} %
shows its effect for the doubly magic nucleus $^{56}$Ni, which
resembles the case of Fig.~\ref{fig:HBRPA-Exact} by having its Fermi
level (common for neutrons and protons) within a gap in its
single-nucleon spectrum. In the case of $^{56}$Ni we have no exact
calculation for comparison, but the interpolation is seen to remove a
certainly unphysical cusp from the curve of the groundstate energy as
a function of $G$.

We finish this section with a discussion of the limit $G \to \infty$.
This discussion is restricted to the case of even $N$ and $Z$. It is
not restricted to $A = 2\Omega$. Without loss of generality the
centroid of the single-nucleon spectrum is supposed to vanish. First
consider the case of degenerate single-nucleon levels, that is,
$\epsilon_k = 0$ for all $k$ so that the Hamiltonian~\eqref{eqn:IVPH}
has only the second term, the pairing force. Our $4\Omega$-dimensional
valence space is then equivalent to a $j$-shell with
$j = \Omega - 1/2$. From the formulas in Ref.~\cite{Ne09} one gets in
this case
\begin{equation}\label{eqn:EdFl}
  E = \frac G2 \left( A \left( \frac A4 - \Omega - \frac 32 \right)
    + T(T + 1) \right) \,. 
\end{equation}
Exactly this expression for the lowest eigenvalue of the pairing force
results from the formulas derived by Edmonds and Flowers~\cite{EF52}
by means of group theory. In other words, for degenerate
single-nucleon levels the HB + RPA gives the exact result. Now assume a
spreading of the levels $\epsilon_k$. Due to their vanishing centroid
the first term in the expression~\eqref{eqn:IVPH} is then, in the
$j$-shell analogy, a sum of spherical tensor components of rank higher
than zero. Since its expectation value in the angular momentum zero
ground state of the degenerate case then vanishes, its contribution to
the energy vanishes in the Born approximation, and the leading term in
this contribution in an expansion in powers of $G^{-1}$ is the linear
term. In other words, for a general single-nucleon spectrum the HB +
RPA result converges asymptotically to the exact one in the limit
$G \to \infty$.

\section{\label{sec:Smooth}Smooth terms}

The smooth sum $\tilde E_\text{s.p.}$ of single-nucleon energies in
Eq.~\eqref{eqn:Strut} is calculated separately for neutrons and
protons by a standard third order Strutinsky smoothing~\cite{BDJ72}
with smoothing parameter $\Gamma = 41 A^{-1/3}$~MeV including energies
until approximately $\tilde\lambda + 5 \Gamma$, where $\tilde\lambda$
is the smooth Fermi level. Corresponding to the spitting of the
pairing energy $P$ into a BCS and RPA part discussed in
Sec.~\ref{sec:RPA} we write
\begin{equation}
  \tilde P = \tilde P_\text{BCS} + \tilde P_\text{RPA} \,.
\end{equation}

Here the part $\tilde P_\text{BCS}$ is calculated separately for
neutrons and protons essentially as suggested by
Brack~\textit{et~al.}~\cite{BDJ72}: We relate the paring strength $G$
to a smooth pair gap $\tilde\Delta$ by considering a half-filled
single-nucleon spectrum of $\Omega$ Kramers doublets with equal
distance $1/\tilde g(\tilde\lambda)$, where $g(\tilde\lambda)$ is the
smooth level density at the smooth Fermi level, and by replacing the
sum in the gap equation with an integral. When this integral is
evaluated more accurately than in Ref.~\cite{BDJ72} one arrives at
\begin{equation}
  \tilde\Delta=\frac{\Omega}
    {2\tilde g(\tilde\lambda)\sinh{a}}
  \quad\text{with}\quad
  a=\frac1{\tilde g(\tilde\lambda)G}\,.
\end{equation}
The BCS correlation energy $P_\text{BCS}$ can be expressed by a sum
over the Kramers doublets, which can be approximated in a similar way
by an integral. By evaluating also this integral more accurately than
in Ref.~\cite{BDJ72}, one gets
\begin{equation}
  \tilde P_\text{BCS} = - \frac{\Omega\tilde\Delta}{2\exp{a}} \,,
\end{equation}

A derivation of an expression for a smooth RPA correlation energy can
be based on Eq.~(38) of Ref.~\cite{Ne09}. We treat first the case of
an $n$ or $p$ space and consider again a half-filled single-nucleon
spectrum of $\Omega$ Kramers doublets with equal distance
$1/\tilde g(\tilde\lambda)$. In Eq.~(38) of Ref.~\cite{Ne09} the sum
over Kramers doublets can be replaced with an integral in the
propagators $G_0(\hat P_\tau,\hat P_\tau,\omega)$ etc. with $\hat
P_\tau = \sum_k \hat P_{\tau,k}$. Some mathematics then leads to
\begin{multline}
  \tilde P_\text{RPA,$\tau$} \\
    = \frac{2\tilde\Delta}\pi\int_0^\infty
        \log{\left(\frac1{2a}\log{\frac{\cosh{(x+a)}}{\cosh{(x-a)}}}
        \right)}\cosh{x}\,dx \,.
\end{multline}
In the derivation of this expression the integral in Eq.~(38) of
Ref.~\cite{Ne09} is displaced to the imaginary $\omega$ axis. This is
allowed because $G_0(\hat P_\tau,\hat P_\tau,\omega)$ etc. are
asymptotically proportional to $\omega^{-2}$.

For the $np$ space one can take into account the discussion in
Sec.~IV~A of Ref.~\cite{Ne09} of the case of a half-filled
infinite spectrum of equidistant quartets. It is shown there that in a
very good approximation the RPA correlation energy in the $np$ space
deviates from that of an $n$ or $p$ space only by a term
$|\lambda_n-\lambda_p|$, where $\lambda_\text{$n$ or $p$}$ is the
chemical potential. Totally we then have
\begin{equation}\begin{aligned}
  \tilde P_\text{RPA} &\ = \sum_{n,p,np} \\
   & \frac{2\tilde\Delta}\pi\int_0^\infty
        \log{\left(\frac1{2a}\log{
          \frac{\cosh{(x+a)}}{\cosh{(x-a)}}
        }\right)}\cosh{x}\,dx \\
    & + |\tilde\lambda_n - \tilde\lambda_p]  \,.
\end{aligned}\end{equation}
Here $\tilde\lambda_{np}$ is taken equal to %
$\tilde\lambda_\text{$n$ or $p$}$ for $N$ or $Z$ equal to $A/2$.

\section{\label{sec:Comparison}Comparison with experimental data}
  
\begin{figure}
  \includegraphics[width=\columnwidth,trim=70 25 90 55]{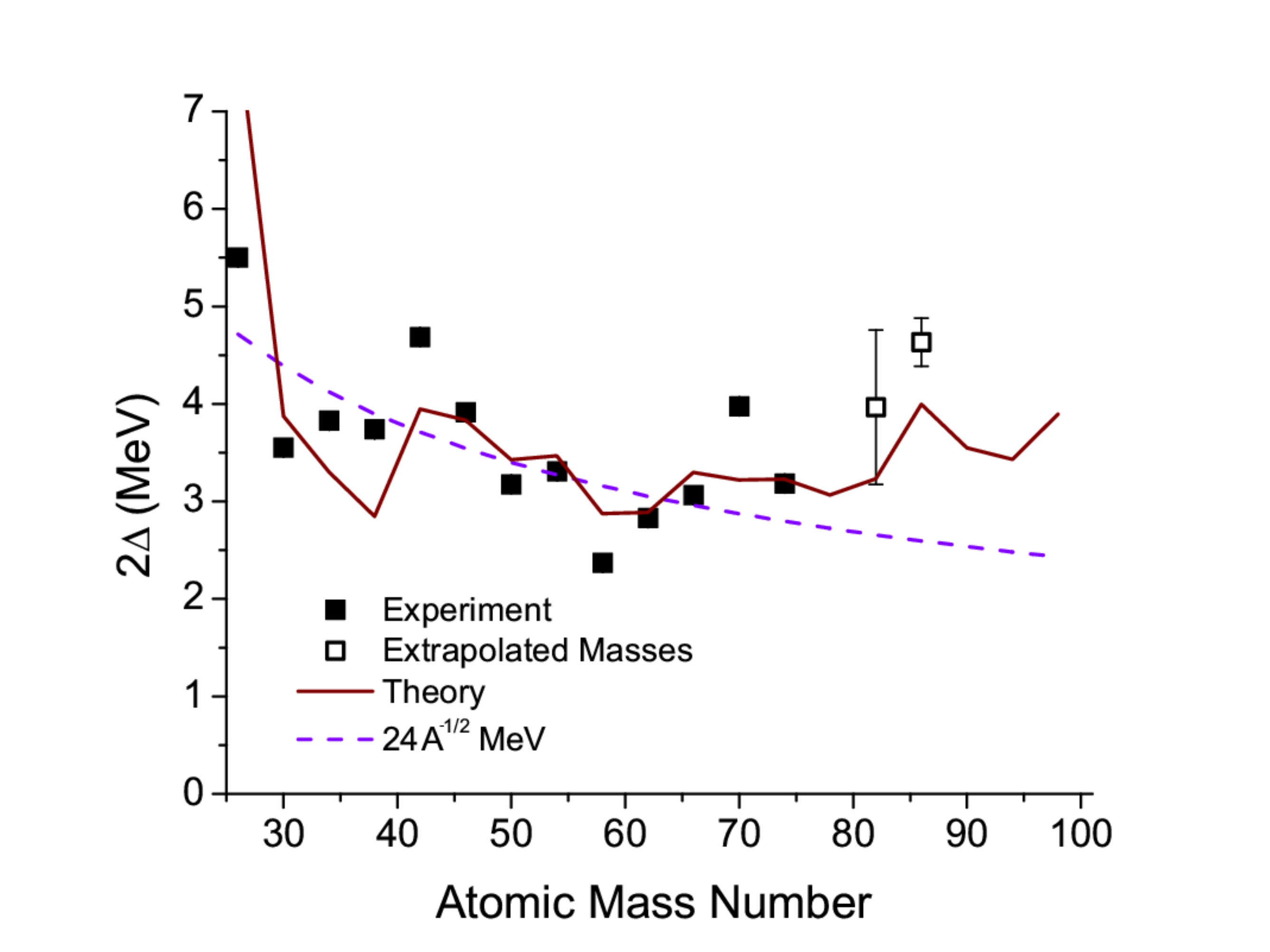}
  \caption{\label{fig:EEOO}%
    (Color online) The $T=0$ even-even-odd-odd binding energy
    difference $2\Delta$ defined by Eq.~\eqref{DeltaEE-OO}. The
    empirical groundstate binding energies are taken from
    Ref.~\cite{Au12} and the $T=0$ excitation energies in the doubly
    odd nuclei from Ref.~\cite{Tu11}. The solid line shows the results
    of our calculations and the purple (gray) dashed line is
    $2\Delta=24 A^{-1/2}$ MeV.}
\end{figure}

As in Ref.~\cite{BF13}, we fit an expression for the pairing force
coupling constant $G$ proportional to a power of $A$ to the empirical
$T=0$ binding energy differences $2\Delta$ of the even-even and
odd-odd $N=Z$ nuclides defined by
\begin{equation}\label{DeltaEE-OO}
   2\Delta = \frac{B(A\!-\!2,0,0)-2B(A,0,0)+B(A\!+\!2,0,0)}2
\end{equation}
for odd $A/2$, cf. the definition of $B(A,T,T_z)$ in the beginning of
Sec.~\ref{sec:BE}. (Note that this $\Delta$ is different from the BCS
pair gap $\Delta$ considered in Secs.~\ref{sec:RPA}
and~\ref{sec:Interpolation}.) We take the groundstate binding
energies from the 2012 Atomic Mass Evaluation and the $T=0$ excitation
energies for odd $A/2$ from the NNDC Evaluated Nuclear Structure Data
Files~\cite{Tu11}. The best fit, shown in Fig.~\ref{fig:EEOO},
is obtained for
\begin{equation}\label{GofA}
  G=8.6 A^{-4/5}\text{ MeV}.
\end{equation}
The calculation is seen to reproduce the observed pattern of
fluctuations due to shell structure of $2\Delta$ as a function of $A$
about a trend line $2\Delta=24 A^{-1/2}$ MeV. For a given $A$, the
calculated and observed values thus lie consistently both above or
both below this line.

\begin{figure}
  \includegraphics[width=\columnwidth,trim=0 20 0 0]{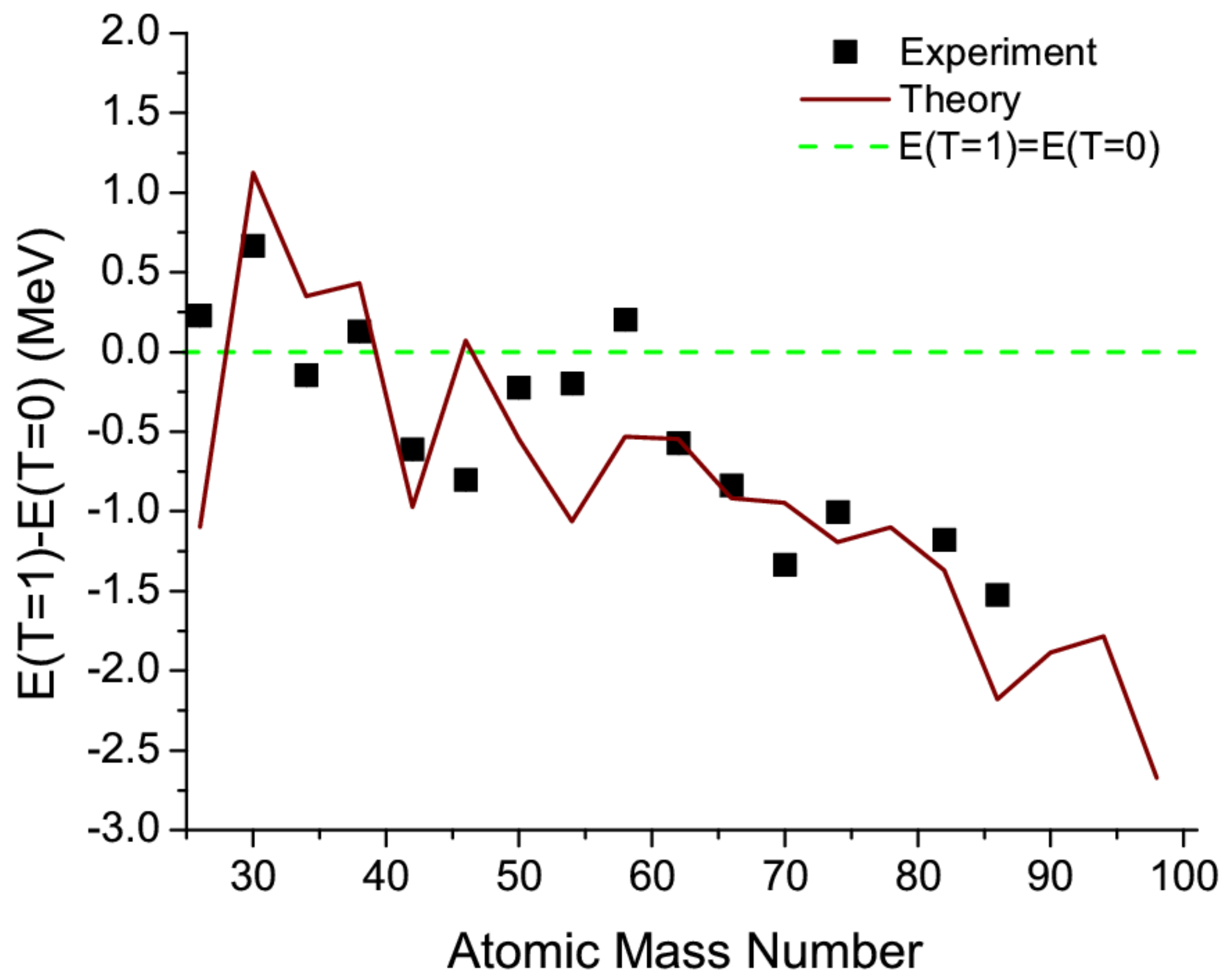}
  \caption{\label{fig:inversion}%
    (Color online) Energy difference of the first $T=1$ and $T=0$
    states in odd-odd $N=Z$ nuclei. The solid line shows the results
    of our calculations and the experimental data are from from
    Ref.~\cite{Tu11}. The green (gray) dashed line is the zero line,
    so when points are below it there is an isospin inversion.}
\end{figure}

\begin{figure}
  \includegraphics[width=\columnwidth,trim=0 20 0 0]{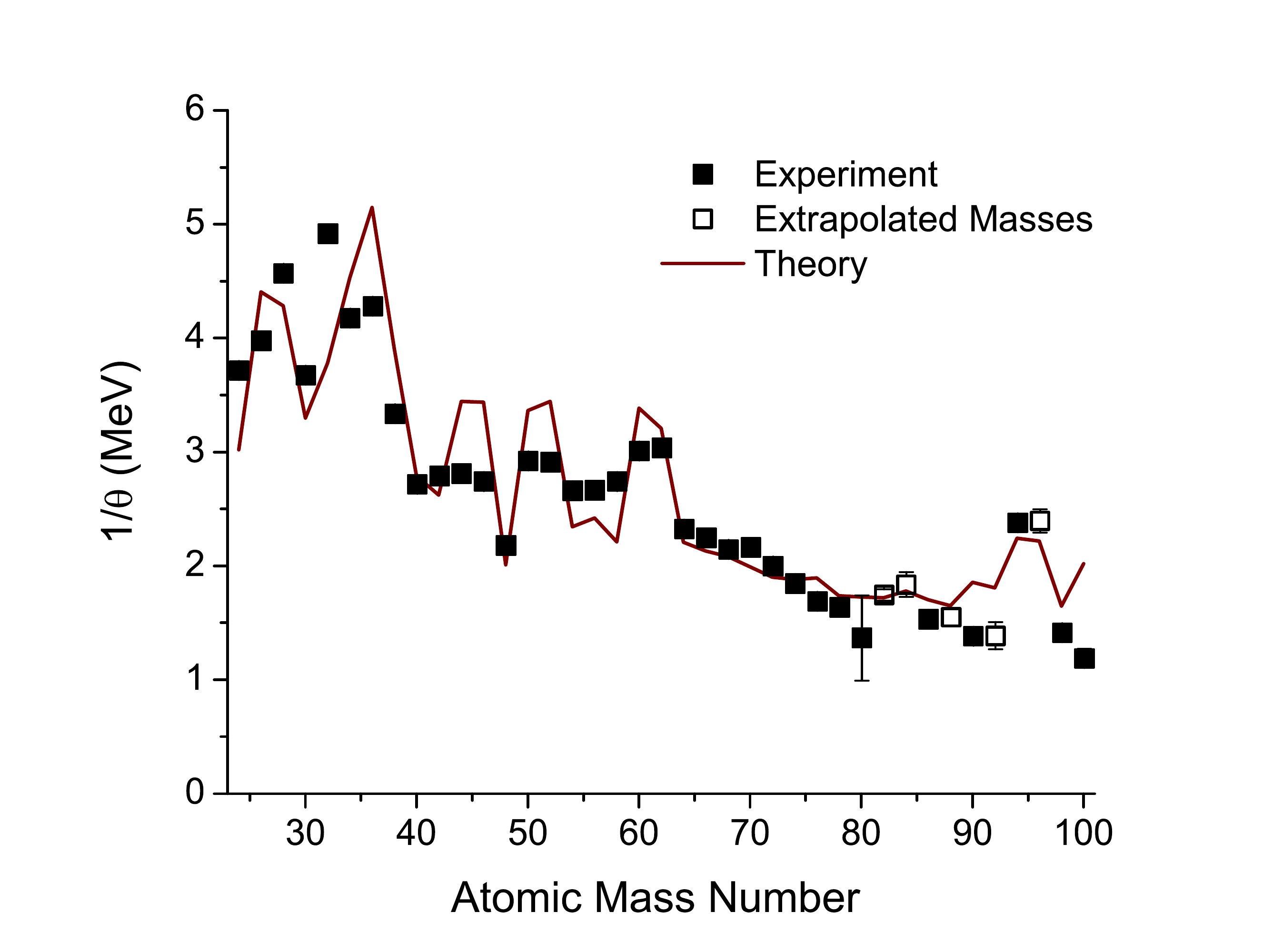}
  \caption{\label{fig:slope}%
    (Color online) The reciprocal isorotational moment of inertia
    $1/\theta$ defined by Eq.~\eqref{eq:isorot}. The empirical values
    are extracted from the binding energies in Ref.~\cite{Au12}. The
    solid line shows the result of our calculations.}
\end{figure}

\begin{figure}
  \includegraphics[width=\columnwidth,trim=0 20 0 0]{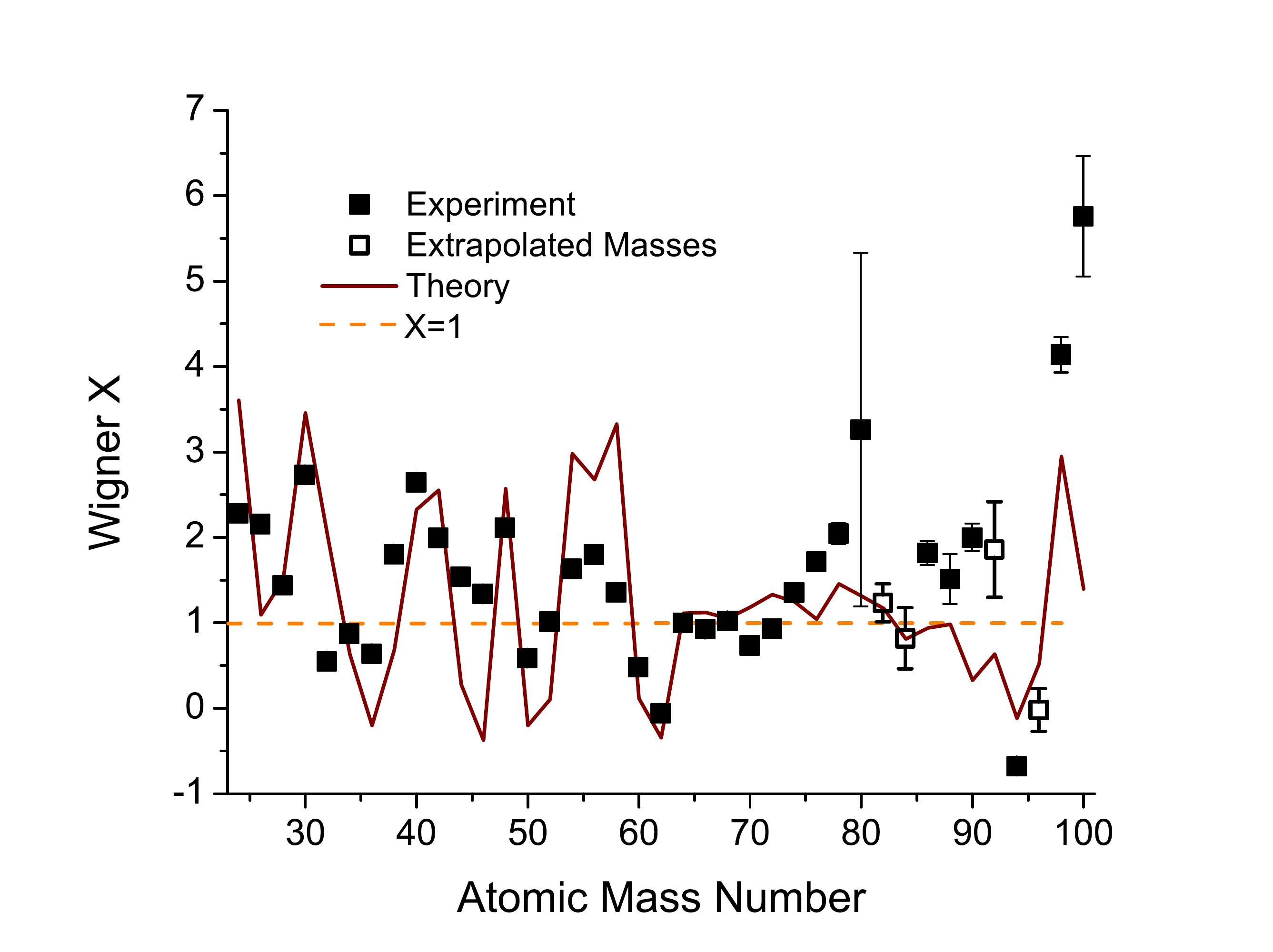}
  \caption{\label{fig:X}%
    (Color online) The Wigner $X$ defined by Eq.~\eqref{eq:isorot}.
    The empirical values are extracted from the binding energies in
    Ref.~\cite{Au12}. The solid line shows the result of our
    calculations. The orange (gray) dashed line is $X=1$.}
\end{figure}

Our further analysis involves Coulomb reduced binding energies. We
assume that the electromagnetic contribution to the total energy is
given by the last term in Eq.~\eqref{eqn:LDM} and thus write for the
remainder
\begin{equation}\label{eq:reduc}
  E_S(A,T) = - B(A,T,T_z) - a_c \frac{Z(Z-1)}{A^{1/3}} B_C \,.
\end{equation}
This reduction is applied to \emph{both} the calculated \emph{and} the
measured binding energies. The error bars shown in Figs.
\ref{fig:EEOO}--\ref{fig:X} include the uncertainties of the empirical
mass differences involved and the uncertainty of $a_c$ in the fit of
Eq. \eqref{eqn:LDM} to the observed masses. When no error bar is
shown, the uncertainty is less than the size of the symbol.

To the extend that the last term in Eq.~\eqref{eq:reduc} may be
assumed to account for all contributions to the total energy from
non-isospin-conserving interactions, $E_S(A,T)$ is independent of
$T_z$. On the right hand side of Eq.~\eqref{eq:reduc}, we mostly
choose $T_z=T$. The only exception is that we compare the difference
$E_S(A,1)-E_S(A,0)$ for odd $A/2$ \emph{measured} for $T_z=0$ with
the one \emph{calculated} for $T_z=T$. 

Figure~\ref{fig:inversion}
shows the measured and calculated values of this difference. The
measured $E_S(A,1)-E_S(A,0)$ is the difference in excitation energy of
the lowest $T=1$ and $T=0$ states of the $N=Z=A/2$ nucleus and is
taken from its Evaluated Nuclear Structure Data File. The observed
trend of a shift from $T = 0$ ground states of the lighter doubly odd
$N = Z$ nuclei in the range $A = $~26--98 to $T = 1$ ground states of
the heavier ones as well as the average slope of the excitation energy
difference as a function of $A$ are well reproduced. As discussed by
Vogel~\cite{Vo00} and Macchiavelli~\textit{et~al.}~\cite{Ma00},
$E_S(A,1)-E_S(A,0)$ can be interpreted as a difference between the
$T = 1$ symmetry energy and the cost in energy $2 \Delta$ of breaking
a Cooper pair. Its downslope as a function of $A$ then results from an
increase of the latter relative to the former. It may be noticed that
$E_S(A,1)-E_S(A,0)$ is the only one of the four combinations of
energies displayed in Figs.~\ref{fig:EEOO}--\ref{fig:X} where measured
and calculated energies with different $T_z$ are compared and the
difference of the energies compared is thus influenced by the Coulomb
reduction~\eqref{eq:reduc}.

We finally consider the quantities $\theta$ and $X$ defined by
\begin{equation}\label{eq:isorot}
  E_S(A,T) = \text{constant} + \frac{T(T+X)}{2\theta} \,,
\end{equation}
where the constant is a function of $A$, and $A/2+T$ is even. The
constant $\theta$ can be interpreted as an isorotational moment of
inertia~\cite{FrSh}, and $X$ is related to the Wigner energy and may
be called the Wigner $X$. In this analysis, $T_z=T$ is chosen
throughout on the right hand side of Eq.~\eqref{eq:reduc}, so only
doubly even nuclei, and therefore only groundstate energies, are
involved. The empirical groundstate binding energies are taken from
the 2012 Atomic Mass Evaluation. The constants $\theta$ and $X$ are
extracted from the calculated and measured $E_S(A,T)$ for $T=0,2,4$ in
the case of even $A/2$ and $T=1,3,5$ in the case of odd $A/2$.

Figures~\ref{fig:slope}
and~\ref{fig:X}
show the result of this analysis. The measured $\theta$ is very well
reproduced including, generally speaking, the features seen in the
experimental values. This is a vast improvement from the calculations
in Ref.~\cite{BF13}, which considerably underestimated $\theta$. This
confirms what was suggested in Ref.~\cite{BF13}, namely that the
underestimate there originated in the small size of the valence space.
For $60<A<80$ the calculation is very accurate, indicating that the
deformations used in the calculations are accurate.

As already mentioned in the introduction, several effects contribute
according to Neerg\aa rd~\cite{Ne09} to the linear term $XT/2\theta$
in the expansion~\eqref{eq:isorot}. One such contribution,
corresponding to $X=1$, comes from the expression $T(T+1)/2\theta$ for
an isorotational energy. This is an average contribution, represented
in our present model by the linear term in the liquid drop symmetry
energy in Eq.~\eqref{eqn:LDM}. Microscopically, a symmetry energy
proportional to $T(T+1)$ is shown in Ref.~\cite{Ne09} to emerge from
the HB + RPA in the idealized case of equidistant single-nucleon
levels. It is dependent in this case on the spontaneous breaking of
the isobaric invariance by the isovector pairing force. In fact, in
the case of equidistant single-nucleon levels the symmetry energy is
proportional to $T(T+X)$ with $X<1$ in the absence of static pair
fields~\cite{Ne03}. The numerical solutions for this case in Fig.~7 of
Ref.\cite{BF13} show that $X$ approaches one from below with
increasing pair coupling strength $G$. Equidistant levels are a
particular favorable case. The same figure demonstrates that $X=1$ is
approached for uneven level distributions as well, because a large $G$
represents the limit of strong deformation in isospace, which results
in rigid isorotation.

On top of this isorotational contribution major contributions to the
linear term in the symmetry energy arise from shell effects. This is
discussed by Neerg\aa rd~\cite{Ne09} and further elaborated by Bentley
and Frauendorf~\cite{BF13}. They relate the deviations of $X$ from one
to the deviation of the distance between the last occupied and first
free level in the absence of pair correlation (cf. Fig.~8 of
Ref.~\cite{BF13}). The realistic pair correlation is too weak to wash
out this consequence of the level bunching. Neerg\aa rd shows, in
particular, that $X$ is large when the $T=0$ Fermi level lies within a
gap in the single-nucleon spectrum. The reason is that the isospin is
then produced by promotion of pairs from proton levels below the gap
to neutron levels above the gap, and each such promotion costs
approximately the same amount of energy equal to twice the gap energy.
That the empirical $X$ is mostly larger than one may thus be seen as
the result of the level density at the $T = 0$ Fermi level being
generally at equilibrium shape lower than corresponding to a uniform
spectrum. By analogy with spatial rotation of nuclei it may be seen as
evidence for a ``softness'' of isorotation. The ratio
$R_{42}=E_{4^+ }/E_{2^+}$, which is 10/3 for a rigid rotor and 2 for a
harmonic vibrator, is commonly used as a measure of how ``rotational''
a nucleus is. The energy of the first few yrast levels of even-even
transitional nuclei can be very well parametrized by the expression
$E(I)=I(I+X)/2{\mathcal J}$. This gives
$X-1 = (10-3R_{42})/(R_{42}-2)$, so $X > 1$ is equivalent to
$2 < R_{42} < 10/3$.

As seen from Fig.~\ref{fig:X}, our model reproduces very well these
fluctuations of $X$ due to shell effects. For $A =$~90, 92, 98, and
100, the calculated valued are markedly below the empirical ones. It
should be noted that the corresponding values of $1/\theta$ in
Fig.~\ref{fig:slope} are markedly above the empirical ones. If one
would take the product of both numbers to obtain $X/\theta$, which is
twice the coefficient of the term linear in $T$ in the
expansion~\eqref{eq:isorot}, a much better agreement would result.

As anticipated from the discussion above, large $X$ occur in the
isobaric chains containing the doubly magic nuclei $^{40}$Ca,
$^{56}$Ni, and $^{100}$Sn. Both chains with odd $A/2$ neighboring each
of these chains with even $A/2$ also have large $X$. This is because
in the odd-$A/2$ chains one pair of neutrons is passive in producing
isospin. This pair just sits for all $T=1,3,5$ in the last quartet
below the gap or the first quartet above the gap while other nucleons
are promoted across both the gap and this quartet repeating the
mechanism described above. A somewhat similar mechanism gives rise to
a large $X$ for $A=30$. There one neutron pair occupies for all
$T=1,3,5$ the $2s_{1/2}$ shell while other nucleons are promoted from
the $1d_{5/2}$ to the $1d_{3/2}$ shell.

For $A=24$ and $A=48$, deformation is involved in producing a large
$X$. In these chains, the $T=0$ nuclei thus have large deformations
while the $T=2$ and $T=4$ nuclei are essentially spherical. Since
isospin is produced by promotion of nucleons within the $1d_{5/2}$ or
$1f_{7/2}$ shell (except that in $^{24}$O one neutron pair occupies the
$2s_{1/2}$ shell), the two spherical nuclei have roughly equal
$E_\text{s.p.}$. We therefore have a large increase of $E_\text{s.p.}$
from $T = 0$ to $T = 2$ due to the departure from the deformed shape
and essentially no such increase from $T = 2$ to $T = 4$. This gives
rise to a large $X$. It also gives a small $1/\theta$ as seen in
Fig.~\ref{fig:slope}.

Tables~\ref{tab:LDparam} and~\ref{tab:LDparam4} show the calculated
and measured binding energies of the individual nuclei and components
of the calculated ones. Also see Note~\cite{supa}. The root mean
square deviation is 0.95~MeV for the 112 doubly even nuclei with a
measured binding energy.

\section{\label{sec:NoIsoscalar}Omission of isoscalar pairing and
  Coulomb interaction}

It has been proposed~\cite{SaWy,Good,Ced,Qi} that the strong
attraction of isoscalar nucleon pairs exhibited by effective shell
model interactions, especially in channels with maximally aligned
nucleonic angular momenta, could give rise, in $N=Z$ nuclei, to a
condensation of such pairs coexisting with or replacing the BCS type
of condensation of isovector pairs. We feel that there are points of
contention with this proposal: (i) The energy of a condensate depends
smoothly on its number of constituents. In the presence of a
condensate of isoscalar pairs the mass of the lowest $T=0$ state of an
$N=Z$ nucleus should therefore depend smoothly on $A$. In reality,
these masses show a staggering with the parity of $A/2$, the doubly
odd masses being elevated above the doubly even ones by an amount
approximately twice the typical BCS pair gap; see Fig.~\ref{fig:EEOO}.
(ii) Bentley and Frauendorf, in their aforesaid study, examine the
effect of adding to the Hamiltonian a schematic interaction of
isoscalar pairs of a neutron and a proton in time-reversed orbits with
an separable structure similar to that of the isovector pairing force.
They find that a weak interaction of this form does not significantly
alter their results, while a stronger one would not allow the model to
reproduce the data. (iii) In the single-$j$-shell model, seniority
zero represents condensation of isovector pairs in the BCS sense.
Neerg\aa rd analyzed the ground states of nuclei with two neutrons and
two protons or two neutron holes and two proton holes in the
$1f_{7/2}$ or $1g_{9/2}$ shell calculated in the single-$j$-shell
approximation with effective interaction from the
literature~\cite{Ne13}. He found these states to have by about 80\%
seniority zero. As pointed out in Neerg\aa rd's study, since the
seniority zero state has a considerable contingent of isoscalar pairs,
the attractive interaction of such pairs stabilizes the seniority zero
component of the state vector rather than competing with it.

For these reasons, we do not consider the possibility of condensation
of isoscalar pairs in our present work. Some studies, for example
Refs.~\cite{MPLV99,Qi}, infer a pairing structure of a shell model
state from counts of nucleon pairs with given angular momentum. Since
Neerg\aa rd demonstrates in Ref.~\cite{Ne13} that such a count is not
a reliable tool for this purpose, we have not taken such work into
account in the discussion in the preceding paragraph.

In Strutinsky-type calculations including schematic isoscalar and
isovector pair correlations, G\l owacz~\textit{et al.}~\cite{Glo04}
achieved results for $E(T=1)-E(T=0)$ in doubly odd $N = Z$ nuclei in
agreement with the data similar to that of our results displayed in
Fig.~\ref{fig:inversion}. This is consistent with the findings of
Bentley and Frauendorf mentioned in the first paragraph of this
section. Although such a mixed scenario cannot be excluded we follow
the principle of Occam's razor, assuming pure isovector pair
correlations.

For $A=48$ we made a supplemental $1f_{7/2}$ shell model calculation
using the interaction ``model~I'' of Zamick and Robinson~\cite{ZR02}
with a normalization to zero of the largest matrix element
(two-nucleon angular momentum $J=6$) so as to make the total
interaction attractive. Like Satu\l a~\textit{et al.}~\cite{SDGMN}, we
switched off successively the interactions in individual $J$ channels.
The results for $\theta$ and $X$ are shown in Table~\ref{table:1f7/2}.
\begin{table}
  \caption{\label{table:1f7/2}
    $1f_{7/2}$ shell model calculation for $A=48$.}
\begin{ruledtabular}
\begin{tabular}{ccccccccccc}
\multicolumn{8}{c}
  {\parbox{.25\columnwidth}{\noindent$J$ of included interactions}}
&\parbox{.15\columnwidth}{$1/\theta$\\(MeV)}
&\parbox{.15\columnwidth}{$X/\theta$\\(MeV)}
&\parbox{.15\columnwidth}{$X$}\\[6pt]
\hline
0&1&2&3&4&5&6&7&2.41&3.17&1.31\\
0& &2&3&4&5&6&7&2.23&2.37&1.07\\
0& &2& &4&5&6&7&1.86&2.29&1.23\\
0& &2& &4& &6&7&1.38&2.42&1.75\\
0& &2& &4& &6& &0.25&0.42&1.71\\
0& &2& &4& & & &0.25&0.42&1.71\\
0& &2& & & & & &0.43&0.61&1.41\\
0& & & & & & & &0.81&0.81&1.00
\end{tabular}
\end{ruledtabular}
\end{table}
Note that because the $J=6$ matrix element is normalized to zero, it
makes no difference whether it is included or not. Like $\theta$ and
$X$ the quantity $\epsilon_W$ displayed in Fig.~2 of Ref.~\cite{SDGMN}
is a function of $E_S(A,T)$ for $T=0,2,4$. The relation is
$\epsilon_W=2X/3\theta$. Like Satu\l a~\textit{et al.}, who include in
their calculations the shells $2p_{3/2}$, $1f_{5/2}$, and $2p_{1/2}$
and employ an interaction appropriate for this larger valence space,
we find that $X/\theta$ and therefore $\epsilon_W$ decreases when the
isoscalar interactions $J=1,3,5,7$ are switched off successively and
very much so when the $J=7$ interaction is switched off finally. It is
seen, however, that this is due \emph{not} to a decrease of $X$, which
actually increases, but to a decrease of the symmetry energy
coefficient $1/2\theta$. The isoscalar shell model interactions thus
contribute significantly to the \emph{entire} symmetry energy and not
just its Wigner term. The reduction of the Wigner energy when the
isoscalar interactions are switched off is only a side effect of this
general reduction of the symmetry energy. The symmetry energy
coefficient reaches its minimum when all the isovector interactions
$J=0,2,4,6$ and none of the isoscalar interactions $J=1,3,5,7$ are
present, and the calculations confirm the well known result derived
analytically by Edmonds and Flowers \cite{EF52} that $X$ is exactly
one for the pure pairing force, $J=0$. In this case $1/\theta = G$ as
seen from Eq.~\eqref{eqn:EdFl}.
   
Our approach makes further simplifying assumptions: (i) The Coulomb
interaction can be treated as a first-order perturbation, that is, its
contribution to the total energy may be approximated by its
expectation value with the wave function determined by the strong
interaction only. This contribution can then be incorporated in the
form of the Coulomb term in the smooth liquid drop binding energy
formula (or subtracted from the experimental binding energies, as done
in this paper). (ii) Isospin-breaking terms of the strong interaction
Hamiltonian (as the difference between the proton and neutron masses)
are neglected. (iii) The difference between proton and neutron mean
fields generates only a constant shift of the proton single-particle
levels relative to the neutron ones. This constant shift drops out in
the shell correction procedure so that one can assume the same single
proton and neutron energies from the outset. These assumptions lead to
our isospin-invariant Hamiltonian (\ref{eqn:IVPH}) used to calculate
shell and pairing corrections.

Sato {\em et al.} \cite{Sato13} studied $N=Z$ nuclei in the framework
of the density functional mean field theory. They describe the $T>0$
states by isocranking about an axis in isospace that is tilted with
respect to the $z$ axis. The resulting quasiparticles are mixtures of
proton and neutron particles and holes. As discussed by Frauendorf and
Sheikh \cite{FrSh}, for an isospin invariant Hamiltonian all mean
field solutions that correspond to the same cranking frequency but a
different orientation of the cranking axis (the ``semicircle'' of
Ref.~\cite{Sato13}) have the same energy. They can be generated by
rotation in isospace from the solution obtained by cranking about the
$z$ axis, which has pure proton and pure neutron quasiparticles. The
rotation generates a mixing of the proton and neutron quasiparticles
(cf. the example of isocranking about the $y$ axis discussed in
Ref.~\cite{FrSh}). Hence if our assumptions hold, it is sufficient to
study isosrotation about the $z$ axis, which generates $T_z=T$
solutions and avoids proton-neutron mixing. The $T_z<T$ solutions are
given by rotation in isospace. If the mean field theory includes the
Coulomb interaction, as the study by Sato~{\em et al.}~\cite{Sato13}
does, the different orientations of the cranking axis are no longer
equivalent, and the orientation of minimal energy has to be
calculated. However the finding of Sato~{\em et al.} that such
solutions lie with a good accuracy on a shifted semicircle indicates
that our assumptions are good approximations.

\section{\label{sec:Sum}Summary and outlook}

A model with nucleons in a charge independent potential well
interacting by an isovector pairing force has been discussed. For a
24-dimensional valence space, the Hartree-Bogolyubov (HB) plus random
phase approximation (RPA) to the lowest eigenvalue of the Hamiltonian
was shown to be accurate except near the values $G_\text{crit.}$ of
the pairing force coupling constant $G$ where the HB solution shifts
from a zero to a non-zero pair gap. The HB + RPA was shown to be
asymptotically exact in the limit $G\to\infty$. To remedy the
inaccuracy of the HB + RPA in the critical regions of $G$ we devised a
scheme of interpolation across the these regions. It is described in
Sec.~\ref{sec:Interpolation}.

The resulting algorithm was used to calculate with a valence space of
dimension twice the mass number $A$ pairing corrections in the
framework of a Nilsson-Strutinsky calculation. For this purpose we
derived in Sec.~\ref{sec:Smooth} expressions for smooth counterterms
to the Bardeen-Cooper-Schrieffer (BCS) and RPA parts of the pair
correlation energy. The deformations and corresponding single-nucleon
energies for the Nilsson-Strutinsky calculation were taken from a
previous Nilsson-Strutinsky plus BCS calculation with the code
\textsc{tac}~\cite{Fr93}. To enforce charge independence the average
of the calculated single-neutron and single-proton energies was
employed. Our expression~\eqref{eqn:LDM} for the macroscopic liquid
drop energy was taken from the work of Duflo and Zuker~\cite{DZ95}
with the omission of a phenomenological pairing energy and has
symmetry energy terms proportional to $T(T+1)$, where $T$ is the
isospin. Its five parameters were fit to the empirical masses
according to the 2012 Atomic Mass Evaluation~\cite{Au12} of the 112
doubly even nuclei with a measured binding energy considered in the
present study.

In this model we calculated the binding energies of the ground states
of the doubly even nuclei with $24 \le A \le 100$ and
$0 \le N-Z \le 10$ and the lowest isospin $T=0$ states of the doubly
odd nuclei with $26 \le A \le 98$ and $N=Z$, where $N$ and $Z$ are the
numbers of neutrons and protons. These calculated binding energies
were compared to the empirical ones from the 2012 Atomic Mass
Evaluation with $T=0$ excitation energies from the NNDC Evaluated
Nuclear Structure Data Files~\cite{Tu11}. In terms of both the
calculated and the empirical binding energies $B(A,T,T_z)$, where
$T_z=(N-Z)/2$, a Coulomb reduced energy $E_S(A,T)$ was defined by
Eq.~\eqref{eq:reduc} with $T_z=T$ on the right-hand side with one
exception to be told later. The following combinations were then
extracted and compared: (i) 2$\Delta$ as defined by
Eq.~\eqref{DeltaEE-OO}. (ii) $E_S(A,1)-E_S(A,0)$ for odd $A/2$. In
this case the \emph{measured} $E_S(A,1)$ was defined with $T_z=0$ on
the right hand side of Eq.~\eqref{eq:reduc}. (iii) The constants
$\theta$ and $X$ in the expansion~\eqref{eq:isorot} for all $A$.

Comparisons of the calculated and measured values of these
combinations are shown in Figs.~\ref{fig:EEOO}--\ref{fig:X}. The
expression~\eqref{GofA} adopted for the pairing force coupling
constant $G$ was fit to the empirical 2$\Delta$. The present
enlargement of the valence space resolved an issue with the constant
$\theta$, which was underestimated in the previous exact calculation
by Bentley and Frauendorf with a 28-dimensional valence space. The
fluctuations of $X$ with $A$ were discussed in
Sec.~\ref{sec:Comparison}. They are well understood from the shell
structure. The root mean square deviation of the calculated and
measured binding energies of the 112 doubly even nuclei with a
measures binding energy is 0.95~MeV.

We anticipate a generalization of the present method to the more
realistic case when neutrons and protons move in different potential
wells due to the Coulomb force. The chief obstacle to this
generalization is the RPA calculation in the $np$ space, which will be
more complex because neutron and proton stationary states no longer
form time-reversed pairs. Simplifying approximations might be
warranted in this step of the procedure, however. The result would be
a method for including pairing correlations beyond a mean field
approximation (BCS, Hartree-Fock-Bogolyubov, relativistic mean field)
that would be simple enough to go on top of any state-of-the-art mean
field approach. This would eliminate the need for the phenomenological
Wigner term often employed in present mean field calculations such as,
for example, those of Refs.~\cite{MN95,HFB21,WS3}.

This work was supported by the DoE Grant DE-FG02-95ER4093.

\bibliography{WignerXIRPA}

\begin{thebibliography}{37}
\bibitem [{sup()}]{supa}%
  \BibitemOpen
  \href@noop {} {}\bibinfo {note} {In
  Tables~\ref{tab:LDparam}--\ref{tab:LDparam4}, $\epsilon_2 = \epsilon\cos
  3\gamma$. Extrapolated empirical binding energies are indicated by an
  asterix.}\BibitemShut {Stop}%
\end{thebibliography}

\begin{longtable*}{cc|cc|ccccccc|cc}
\caption{Deformations and Binding Energy Contributions for Even-Even
  Nuclei \label{tab:LDparam}}\\
\toprule $N$ & $Z$ & $\epsilon_2$ & $\epsilon_4$
& $E_\text{DLD}$ & $E_\text{s.p.}\!-\!\tilde E_\text{s.p.}$
& $P_\text{BCS}$ & $\tilde P_\text{BCS}$ & $P_\text{RPA}$
& $\tilde P_\text{RPA}$ & $P\!-\!\tilde P$ & $B_\text{Calc.}$ &
$B_\text{Exp.}$(Error) \cite{Au12}\\
&&&&&& (MeV) & (MeV) & (MeV) & (MeV) & (MeV) & (MeV) & (MeV) \\
\hline\endfirsthead
\toprule $N$ & $Z$ & $\epsilon_2$ & $\epsilon_4$
& $E_\text{DLD}$ & $E_\text{s.p.}\!-\!\tilde E_\text{s.p.}$
& $P_\text{BCS}$ & $\tilde P_\text{BCS}$ & $P_\text{RPA}$
& $\tilde P_\text{RPA}$ & $P\!-\!\tilde P$ & $B_\text{Calc.}$ &
$B_\text{Exp.}$(Error) \cite{Au12}\\
&&&&&& (MeV) & (MeV) & (MeV) & (MeV) & (MeV) & (MeV) & (MeV) \\
\hline\endhead
\hline\multicolumn{13}{r}{\textit{Continues}}\endfoot
\hline\endlastfoot
12&12&0.284&0.014&-196.494&-1.599&0.000&-0.208&-15.263&-15.966&0.911&197.182&198.257(0.000)\\
14&10&0.091&0.000&-191.739&2.999&-0.874&-0.254&-13.751&-13.448&-0.923&189.663&191.840(0.001)\\
16&8&0.000&0.000&-162.435&-4.428&0.000&-0.315&-12.576&-10.691&-1.570&168.433&168.952(0.110)\\
14&12&0.201&0.012&-216.049&1.347&-0.358&-0.279&-14.924&-15.121&0.118&214.584&216.681(0.000)\\
16&10&0.000&0.000&-201.290&2.183&-1.966&-0.313&-13.362&-12.708&-2.307&201.414&201.551(0.018)\\
18&8&0.000&0.000&-163.793&-2.082&-1.178&-0.382&-12.341&-10.059&-3.078&168.953&168.862(0.156)\\
14&14&-0.222&-0.003&-233.778&-1.686&0.000&-0.310&-15.841&-16.543&1.012&234.452&236.537(0.000)\\
16&12&0.000&0.000&-231.608&3.086&-2.013&-0.333&-14.332&-14.370&-1.642&230.164&231.627(0.002)\\
18&10&0.000&0.000&-207.636&4.304&-3.107&-0.374&-13.146&-12.042&-3.837&207.169&206.882(0.096)\\
16&14&0.000&0.000&-254.613&-0.669&-0.035&-0.365&-15.489&-15.782&0.623&254.659&255.620(0.000)\\
18&12&0.000&0.000&-242.838&5.142&-3.097&-0.386&-14.000&-13.686&-3.025&240.721&241.635(0.003)\\
20&10&0.000&0.000&-211.426&2.649&-1.803&-0.451&-13.216&-11.451&-3.117&211.894&211.276(0.280)\\
16&16&0.000&0.000&-271.276&-2.921&0.000&-0.399&-16.240&-17.023&1.182&273.015&271.780(0.000)\\
18&14&0.000&0.000&-270.577&1.426&-1.172&-0.411&-15.080&-15.088&-0.753&269.904&271.407(0.000)\\
20&12&0.000&0.000&-251.164&3.507&-1.834&-0.456&-14.041&-13.076&-2.343&250.000&249.723(0.003)\\
18&16&0.000&0.000&-291.799&-0.823&-1.108&-0.438&-15.741&-16.322&-0.089&292.711&291.839(0.000)\\
20&14&0.000&0.000&-283.337&-0.096&-0.016&-0.473&-15.038&-14.466&-0.115&283.548&283.429(0.014)\\
22&12&0.000&0.000&-256.998&7.279&-4.331&-0.557&-13.666&-12.533&-4.907&254.626&256.713(0.029)\\
18&18&0.000&0.000&-307.277&1.193&-2.144&-0.471&-16.301&-17.438&-0.536&306.620&306.717(0.000)\\
20&16&0.000&0.000&-308.862&-2.273&0.000&-0.494&-15.640&-15.694&0.548&310.587&308.714(0.000)\\
22&14&0.000&0.000&-293.328&3.674&-2.546&-0.566&-14.649&-13.912&-2.717&292.371&292.008(0.071)\\
20&18&0.000&0.000&-328.496&-0.268&-1.041&-0.521&-16.290&-16.806&-0.004&328.768&327.343(0.000)\\
22&16&0.000&0.000&-322.915&1.468&-2.503&-0.580&-15.225&-15.133&-2.015&323.462&321.054(0.007)\\
24&14&0.132&-0.005&-300.467&0.803&-0.674&-0.683&-14.796&-13.436&-1.351&301.015&299.928(0.070)\\
20&20&0.000&0.000&-342.868&-1.679&0.000&-0.565&-17.132&-17.828&1.261&343.286&342.052(0.000)\\
22&18&0.000&0.000&-346.496&3.375&-3.474&-0.600&-15.784&-16.242&-2.416&345.537&343.810(0.000)\\
24&16&0.000&0.000&-334.329&2.113&-3.206&-0.703&-14.997&-14.634&-2.866&335.082&333.173(0.004)\\
22&20&0.000&0.000&-364.689&1.929&-2.409&-0.637&-16.675&-17.264&-1.183&363.943&361.896(0.000)\\
24&18&0.000&0.000&-361.658&3.980&-4.121&-0.716&-15.506&-15.739&-3.172&360.850&359.336(0.006)\\
26&16&0.000&0.000&-343.414&-0.002&-2.383&-0.862&-15.000&-14.188&-2.333&345.749&344.116(0.003)\\
22&22&0.000&0.000&-378.017&5.424&-4.720&-0.701&-16.899&-18.207&-2.711&375.304&375.475(0.001)\\
24&20&0.000&0.000&-383.498&2.548&-3.088&-0.745&-16.292&-16.759&-1.876&382.826&380.960(0.000)\\
26&18&0.000&0.000&-374.304&1.869&-3.295&-0.868&-15.502&-15.288&-2.641&375.076&373.729(0.002)\\
24&22&0.000&0.000&-400.365&5.983&-5.354&-0.800&-16.595&-17.702&-3.447&397.829&398.196(0.000)\\
26&20&0.000&0.000&-399.626&0.490&-2.316&-0.889&-16.242&-16.305&-1.364&400.500&398.772(0.002)\\
28&18&0.000&0.000&-384.707&-2.675&-0.920&-1.047&-15.575&-14.881&-0.567&387.949&386.929(0.041)\\
24&24&0.150&-0.014&-411.947&-0.464&-0.768&-0.862&-18.238&-18.561&0.417&411.994&411.469(0.007)\\
26&22&0.000&0.000&-419.885&3.908&-4.566&-0.936&-16.599&-17.247&-2.982&418.959&418.703(0.000)\\
28&20&0.000&0.000&-413.352&-3.969&0.000&-1.061&-16.281&-15.893&0.673&416.648&416.001(0.000)\\
26&24&0.100&-0.002&-435.206&0.153&-1.545&-0.997&-17.592&-18.108&-0.032&435.085&435.051(0.001)\\
28&22&0.000&0.000&-436.862&-0.538&-2.241&-1.100&-16.670&-16.833&-0.978&438.378&437.785(0.000)\\
30&20&0.000&0.000&-424.920&-1.037&-1.451&-1.249&-16.231&-15.517&-0.916&426.873&427.508(0.002)\\
26&26&0.000&0.000&-446.900&2.481&-4.416&-1.134&-17.820&-18.915&-2.187&446.606&447.700(0.007)\\
28&24&0.000&0.000&-455.667&0.084&-2.878&-1.172&-17.139&-17.701&-1.144&456.727&456.350(0.001)\\
30&22&0.000&0.000&-451.546&2.310&-3.602&-1.280&-16.545&-16.453&-2.414&451.650&451.966(0.007)\\
28&26&0.000&0.000&-470.139&-1.832&-2.155&-1.280&-17.881&-18.500&-0.256&472.227&471.764(0.000)\\
30&24&0.000&0.000&-473.400&2.888&-4.205&-1.344&-16.977&-17.319&-2.519&473.031&474.008(0.001)\\
32&22&0.000&0.000&-464.150&2.970&-3.055&-1.466&-16.399&-16.102&-1.886&463.066&464.237(0.125)\\
28&28&0.000&0.000&-480.607&-6.040&0.000&-1.416&-18.436&-19.234&2.214&484.433&483.995(0.001)\\
30&26&0.000&0.000&-490.853&0.962&-3.480&-1.443&-17.634&-18.117&-1.554&491.445&492.259(0.000)\\
32&24&0.000&0.000&-488.938&3.533&-3.662&-1.523&-16.813&-16.965&-1.987&487.392&488.499(0.002)\\
30&28&0.000&0.000&-504.227&-3.230&-1.350&-1.571&-18.371&-18.851&0.701&506.756&506.459(0.000)\\
32&26&0.000&0.000&-509.265&1.623&-2.963&-1.613&-17.428&-17.761&-1.017&508.659&509.950(0.000)\\
34&24&0.087&0.002&-502.230&2.988&-2.722&-1.706&-16.779&-16.626&-1.169&500.411&501.195(0.203)\\
30&30&0.000&0.000&-513.812&-0.485&-2.653&-1.717&-18.900&-19.528&-0.308&514.605&514.982(0.001)\\
32&28&0.000&0.000&-525.451&-2.529&-0.889&-1.733&-18.075&-18.493&1.262&526.718&526.846(0.000)\\
34&26&0.000&0.000&-525.572&4.579&-4.999&-1.784&-17.159&-17.428&-2.946&523.939&525.350(0.003)\\
32&30&0.000&0.000&-537.780&0.177&-2.182&-1.871&-18.621&-19.170&0.238&537.365&538.119(0.001)\\
34&28&0.000&0.000&-544.476&0.440&-2.958&-1.896&-17.762&-18.159&-0.665&544.701&545.262(0.000)\\
36&26&-0.043&0.001&-539.889&4.484&-4.551&-1.961&-17.011&-17.114&-2.487&537.892&538.959(0.003)\\
32&32&0.000&0.000&-546.508&0.826&-1.732&-2.017&-19.131&-19.799&0.953&544.729&545.845(0.004)\\
34&30&0.000&0.000&-559.467&3.087&-4.221&-2.026&-18.291&-18.834&-1.652&558.032&559.098(0.001)\\
36&28&0.000&0.000&-561.482&1.511&-3.579&-2.060&-17.540&-17.844&-1.215&561.186&561.757(0.001)\\
34&32&0.091&0.004&-570.523&2.342&-2.801&-2.173&-18.838&-19.464&-0.002&568.183&569.279(0.002)\\
36&30&-0.037&0.001&-579.006&3.232&-3.995&-2.189&-18.078&-18.519&-1.365&577.139&578.136(0.001)\\
38&28&0.000&0.000&-576.625&0.810&-2.899&-2.231&-17.420&-17.548&-0.540&576.355&576.808(0.001)\\
34&34&-0.171&-0.002&-577.748&1.959&-2.569&-2.277&-19.364&-20.050&0.394&575.395&576.439(0.000)\\
36&32&-0.113&0.002&-592.486&3.372&-3.622&-2.323&-18.566&-19.145&-0.720&589.834&590.793(0.002)\\
38&30&0.000&0.000&-596.694&3.400&-4.112&-2.349&-17.911&-18.221&-1.453&594.747&595.386(0.001)\\
36&34&-0.213&-0.002&-601.773&1.828&-2.874&-2.428&-19.205&-19.731&0.080&599.865&600.322(0.002)\\
38&32&-0.121&0.005&-612.512&3.318&-3.691&-2.506&-18.403&-18.848&-0.740&609.934&610.519(0.001)\\
40&30&0.000&0.000&-612.537&2.383&-3.551&-2.532&-17.808&-17.939&-0.888&611.042&611.086(0.002)\\
36&36&-0.273&-0.003&-607.918&-0.804&-1.434&-2.570&-19.937&-20.289&1.488&607.234&606.911(0.008)\\
38&34&-0.200&0.002&-624.444&2.469&-3.375&-2.646&-18.947&-19.436&-0.240&622.215&622.403(0.002)\\
40&32&0.000&0.000&-631.231&2.986&-3.122&-2.653&-18.215&-18.564&-0.120&628.365&628.686(0.000)\\
38&36&-0.248&0.001&-633.186&1.431&-3.102&-2.802&-19.574&-19.994&0.120&631.635&631.445(0.002)\\
40&34&-0.190&0.008&-645.134&1.908&-3.019&-2.866&-18.820&-19.156&0.183&643.043&642.891(0.000)\\
42&32&0.000&0.000&-647.719&4.019&-4.995&-2.859&-18.078&-18.296&-1.918&645.618&645.665(0.000)\\
38&38&-0.238&0.006&-639.705&2.508&-4.048&-3.007&-19.917&-20.524&-0.434&637.631&637.939(0.034)\\
40&36&-0.220&0.008&-656.480&2.144&-3.574&-3.035&-19.272&-19.715&-0.096&654.432&654.270(0.004)\\
42&34&0.000&0.000&-665.112&6.732&-6.951&-2.978&-18.393&-18.882&-3.484&661.864&662.072(0.000)\\
40&38&-0.218&0.013&-665.086&2.461&-4.005&-3.224&-19.695&-20.244&-0.232&662.857&663.007(0.007)\\
42&36&-0.201&0.014&-677.723&2.205&-3.861&-3.268&-19.114&-19.451&-0.256&675.774&675.578(0.001)\\
44&34&0.058&0.000&-682.075&3.823&-5.372&-3.246&-18.361&-18.633&-1.854&680.106&679.989(0.000)\\
40&40&-0.212&0.020&-670.705&1.787&-3.677&-3.427&-20.231&-20.743&0.262&668.656&669.929(1.490)\\
42&38&-0.205&0.018&-688.423&2.006&-3.898&-3.450&-19.556&-19.978&-0.026&686.443&686.288(0.003)\\
44&36&0.063&0.001&-698.213&4.544&-5.757&-3.375&-18.704&-19.186&-1.900&695.569&695.434(0.001)\\
42&40&-0.204&0.025&-696.104&1.021&-3.372&-3.646&-20.086&-20.478&0.666&694.417&694.458(0.164)*\\
44&38&-0.073&0.003&-711.058&4.148&-5.495&-3.525&-19.101&-19.713&-1.358&708.268&708.129(0.006)\\
46&36&0.051&0.002&-715.895&2.810&-5.457&-3.637&-18.703&-18.943&-1.580&714.665&714.274(0.001)\\
42&42&-0.200&0.031&-700.894&0.124&-3.045&-3.861&-20.630&-20.952&1.138&699.632&699.636(0.420)*\\
44&40&0.000&0.000&-721.028&5.488&-7.106&-3.607&-19.427&-20.208&-2.718&718.258&718.117(0.006)\\
46&38&0.000&0.000&-730.971&4.551&-7.171&-3.733&-19.009&-19.463&-2.984&729.404&728.911(0.001)\\
44&42&0.000&0.000&-727.856&6.454&-8.833&-3.781&-19.771&-20.688&-4.135&725.537&725.385(0.004)\\
46&40&0.000&0.000&-742.815&3.596&-6.638&-3.873&-19.389&-19.965&-2.189&741.408&740.808(0.004)\\
48&38&0.000&0.000&-749.119&1.310&-5.577&-4.047&-19.169&-19.233&-1.466&749.275&748.927(0.001)\\
44&44&0.000&0.000&-731.868&5.965&-9.440&-3.992&-20.144&-21.146&-4.446&730.349&730.224(0.264)*\\
46&42&0.000&0.000&-751.670&4.570&-8.370&-4.041&-19.729&-20.444&-3.614&750.714&750.104(0.004)\\
48&40&0.000&0.000&-762.974&0.388&-5.074&-4.182&-19.546&-19.734&-0.704&763.290&762.610(0.005)\\
46&44&0.000&0.000&-757.674&4.099&-8.991&-4.247&-20.106&-20.902&-3.948&757.523&756.879(0.004)\\
48&42&0.000&0.000&-773.811&1.378&-6.820&-4.345&-19.887&-20.213&-2.149&774.582&773.733(0.004)\\
50&40&0.000&0.000&-781.602&-4.065&-2.113&-4.529&-19.296&-19.514&2.634&783.033&783.898(0.002)\\
46&46&0.000&0.000&-760.954&2.262&-8.560&-4.495&-20.605&-21.339&-3.331&762.023&761.668(0.460)*\\
48&44&0.000&0.000&-781.766&0.934&-7.460&-4.545&-20.268&-20.670&-2.513&783.345&782.439(0.003)\\
50&42&0.000&0.000&-794.376&-3.049&-3.878&-4.686&-19.645&-19.993&1.156&796.269&796.510(0.001)\\
48&46&0.000&0.000&-786.965&-0.867&-7.052&-4.787&-20.771&-21.107&-1.929&789.761&788.817(0.004)\\
50&44&0.000&0.000&-804.240&-3.458&-4.545&-4.881&-20.038&-20.450&0.748&806.950&806.864(0.003)\\
52&42&0.000&0.000&-813.453&1.390&-7.108&-5.056&-19.673&-19.783&-1.942&814.005&814.258(0.000)\\
48&48&0.000&0.000&-789.526&-3.952&-5.585&-5.074&-21.534&-21.523&-0.522&794.000&792.864(0.384)*\\
50&46&0.000&0.000&-811.319&-5.216&-4.180&-5.118&-20.571&-20.887&1.254&815.281&815.041(0.004)\\
52&44&0.000&0.000&-825.186&0.953&-7.762&-5.246&-20.028&-20.240&-2.304&826.537&826.502(0.000)\\
50&48&0.000&0.000&-815.732&-8.249&-2.763&-5.399&-21.388&-21.303&2.551&821.430&821.073(0.052)\\
52&46&0.000&0.000&-834.107&-0.824&-7.395&-5.478&-20.498&-20.676&-1.739&836.670&836.322(0.005)\\
54&44&0.000&0.000&-844.686&4.129&-9.659&-5.625&-19.618&-20.039&-3.613&844.170&844.791(0.006)\\
50&50&0.000&0.000&-817.586&-12.489&0.000&-5.718&-21.073&-21.699&6.344&823.731&825.297(0.302)\\
52&48&0.000&0.000&-840.335&-3.866&-5.985&-5.753&-21.186&-21.092&-0.326&844.527&843.774(0.002)\\
54&46&0.000&0.000&-855.412&2.343&-9.305&-5.852&-20.060&-20.475&-3.038&856.107&856.371(0.018)
\end{longtable*}

\begin{longtable*}{cc|cc|ccccccc|cc}
\caption{Deformations and  Contributions to the $T=0$ Binding Energy
  for Odd-Odd Nuclei \label{tab:LDparam4}}\\
\toprule $N$ & $Z$ & $\epsilon_2$ & $\epsilon_4$
& $E_\text{DLD}$ & $E_\text{s.p.}\!-\!\tilde E_\text{s.p.}$
& $P_\text{BCS}$ & $\tilde P_\text{BCS}$ & $P_\text{RPA}$
& $\tilde P_\text{RPA}$ & $P\!-\!\tilde P$ & $B_\text{Calc.}$ &
$B_\text{Exp.}$(Error) \cite{Au12}\\
&&&&&& (MeV) & (MeV) & (MeV) & (MeV) & (MeV) & (MeV) & (MeV) \\
\hline\endfirsthead
\toprule $N$ & $Z$ & $\epsilon_2$ & $\epsilon_4$
& $E_\text{DLD}$ & $E_\text{s.p.}\!-\!\tilde E_\text{s.p.}$
& $P_\text{BCS}$ & $\tilde P_\text{BCS}$ & $P_\text{RPA}$
& $\tilde P_\text{RPA}$ & $P\!-\!\tilde P$ & $B_\text{Calc.}$ &
$B_\text{Exp.}$(Error) \cite{Au12}\\
&&&&&& (MeV) & (MeV) & (MeV) & (MeV) & (MeV) & (MeV) & (MeV) \\
\hline\endhead
\hline\multicolumn{13}{r}{\textit{Continues}}\endfoot
\hline\endlastfoot
13&13&0.031&0.006&-216.587&6.083&0.000&-0.281&-14.087&-16.296&2.490&208.014&211.894(0.000)\\
15&15&-0.111&-0.002&-252.841&0.034&0.000&-0.365&-14.224&-16.801&2.942&249.865&250.605(0.000)\\
17&17&0.000&0.000&-289.325&0.086&0.000&-0.434&-14.951&-17.235&2.718&286.521&285.419(0.000)\\
19&19&0.000&0.000&-325.126&0.559&0.000&-0.514&-15.694&-17.635&2.455&322.112&320.646(0.000)\\
21&21&0.000&0.000&-360.499&2.682&0.000&-0.626&-16.181&-18.019&2.464&355.353&354.076(0.000)\\
23&23&0.075&-0.007&-395.239&3.580&-0.673&-0.784&-16.664&-18.389&1.836&389.823&389.560(0.000)\\
25&25&0.075&-0.007&-429.677&1.752&-0.440&-0.994&-17.252&-18.742&2.044&425.881&426.409(0.001)\\
27&27&0.000&0.000&-463.816&-1.258&0.000&-1.272&-17.336&-19.078&3.014&462.060&462.540(0.000)\\
29&29&0.000&0.000&-497.273&-2.700&0.000&-1.566&-17.626&-19.384&3.324&496.649&497.116(0.001)\\
31&31&0.000&0.000&-530.224&0.671&-0.326&-1.868&-18.442&-19.666&2.766&526.787&527.584(0.001)\\
33&33&-0.086&-0.001&-562.430&3.069&-0.678&-2.159&-18.818&-19.927&2.590&556.771&558.078(0.006)\\
35&35&-0.222&-0.003&-592.989&1.898&-0.470&-2.414&-19.125&-20.171&2.990&588.101&587.700(0.015)\\
37&37&-0.256&0.002&-623.894&1.593&-0.574&-2.789&-19.534&-20.409&3.090&619.211&619.241(0.003)\\
39&39&-0.225&0.013&-655.291&2.738&-1.475&-3.218&-19.911&-20.635&2.467&650.086&\empt\\
41&41&-0.206&0.026&-685.874&1.341&-0.875&-3.642&-20.002&-20.849&3.614&680.919&680.814(0.328)*\\
43&43&-0.100&0.016&-716.766&3.893&-3.156&-3.896&-19.915&-21.049&1.874&710.999&710.298(0.258)*\\
45&45&0.000&0.000&-746.475&4.439&-6.121&-4.234&-19.961&-21.243&-0.605&742.641&\empt\\
47&47&0.000&0.000&-775.304&-0.543&-4.287&-4.775&-20.659&-21.432&1.261&774.586&\empt\\
49&49&0.000&0.000&-803.620&-7.941&0.000&-5.389&-20.418&-21.613&6.584&804.977&\empt
\end{longtable*}

\end{document}